\documentclass[runningheads]{llncs}
\usepackage{graphicx}
\usepackage[margin=1.4in]{geometry}

\usepackage{booktabs} % For formal tables
\usepackage{amsmath,amsfonts,amssymb,graphicx,indentfirst,bm,boldline}
\usepackage[numbers]{natbib}
\usepackage{float}
\usepackage[toc,page]{appendix}

\usepackage{amsmath,graphicx,indentfirst,bm,boldline}
\usepackage{dsfont}
\newcommand{\indicator}[1]{\textbf{1}_{\left[{#1}\right]}}
\newcommand{\shadingFunc}{\beta}
\newcommand{\Exp}[1]{\mathbb{E}\left(#1\right)}
\newcommand{\Expb}[1]{\mathbb{E}\left(#1\right)}

\newcommand{\eps}{\epsilon}
\newcommand{\redtext}[1]{\textcolor{red}{#1}}
\newcommand{\tendsto}{\rightarrow}

\newcommand{\trsp}{'}
\newcommand{\newThreshNotation}{\tilde{\beta}^{(\epsilon)}_r}

\newtheorem{innercustomthm}{Theorem}
\newenvironment{customthm}[1]
  {\renewcommand\theinnercustomthm{#1}\innercustomthm}
  {\endinnercustomthm}
  
  \newtheorem{innercustomlemma}{Lemma}
\newenvironment{customlemma}[1]
  {\renewcommand\theinnercustomlemma{#1}\innercustomlemma}
  {\endinnercustomlemma}
\newcommand{\newThreshNotationEpsn}{\tilde{\beta}^{(\epsilon_n)}_r}

\newcommand{\vValue}{\psi}

\DeclareMathOperator*{\argmax}{\textrm{argmax}}
\usepackage{hyperref}
\usepackage{multirow}
\usepackage{stfloats}
\usepackage{appendix}
\usepackage{tikz}
\usetikzlibrary{positioning}

\setcitestyle{acmnumeric}

\begin{document}
\title{Thresholding at the monopoly price: an agnostic way to improve bidding strategies in revenue-maximizing auctions}
\titlerunning{Thresholding at the monopoly price}
% If the paper title is too long for the running head, you can set
% an abbreviated paper title here
%
\author{Thomas Nedelec\inst{1,2} \and
Clément Calauzènes\inst{1} \and
Vianney Perchet\inst{1,3}\and 
Noureddine El Karoui \inst{1,4}}
\authorrunning{T. Nedelec et al.}
% First names are abbreviated in the running head.
% If there are more than two authors, 'et al.' is used.
%
\institute{Criteo AI Lab \and
ENS Paris Saclay \and 
ENSAE \and
UC, Berkeley
}
%Springer Heidelberg, Tiergartenstr. 17, 69121 Heidelberg, Germany
%\email{lncs@springer.com}\\
%\url{http://www.springer.com/gp/computer-science/lncs} \and
%ABC Institute, Rupert-Karls-University Heidelberg, Heidelberg, Germany\\
%\email{\{abc,lncs\}@uni-heidelberg.de}}
%
\maketitle              % typeset the header of the contribution
\begin{abstract}
We address the problem of improving bidders' strategies in prior-dependent revenue-maximizing auctions and introduce a simple and generic method to design novel bidding strategies if the seller uses past bids to optimize her mechanism. We propose a simple and agnostic strategy, independent of the distribution of the competition, that is robust to mechanism changes and local (as opposed to global) optimization of e.g. reserve prices by the seller. This strategy guarantees an increase in utility compared to the truthful strategy for any distribution of the competition. In textbook-style examples, for instance with uniform [0,1] value distributions and two bidders, this no-side-information and mechanism-independent strategy yields an enormous 57\% increase in buyer utility for lazy second price auctions with monopoly reserves. 

When the bidder knows the distribution of the highest bid of the competition, we show how to optimize the tradeoff between reducing the reserve price and beating the competition. Our formulation enables to study some important robustness properties of the strategies, showing their impact even when the seller is using a data-driven approach to set the reserve prices. In this sample-size setting, we prove under what conditions, thresholding bidding strategies can still improve the buyer’s utility.

The gist of our approach is to see optimal auctions in practice as a Stackelberg game where the buyer is the leader, as he is the first one to move (here bid) when the seller is the follower as she has no prior information on the bidder. 
\keywords{Revenue-maximizing auctions \and strategic bidders \and reserve prices.}
\end{abstract}
\section{Introduction}
Auctions currently play a key role in the internet ecosystem through online advertising \citep{AllouahBesbes2017,Balseiro2015MultiStage,AmiKeaKey12}. Ad slots are sold to  advertisers by a publisher following more or less explicit mechanisms, i.e., a type of auction with specific rules. Those auctions take place on platforms called ``ad exchanges'' \citep{Mut09}, either in a multi-item fashion in the search setting or in a single-item fashion in the display advertising setting. One of the most common type of auctions used in the latter setting is the classical second price auctions with  reserve prices. These auctions are myopically truthful, as it is dominant for buyers to bid their true valuation at each individual auction, and even revenue-maximizing for identical bidders \citep{Myerson81,RilSam81}.

Since the influential paper of Myerson \cite{Myerson81}, the revenue-maximizing auction literature has assumed that bidders' value distributions  are common knowledge among the seller and the bidders.  The classical reasoning \citep{AllouahBesbes2017,Golrezaei2017} behind this traditional assumption is that the seller is choosing incentive compatible auctions such as Vickrey auctions; ``hence'', since in a one shot second price auction it is optimal to bid one's own valuations, the seller can safely expect that past bids reflect past valuations. An approximation of the buyer's distribution of valuations easily follows. Then, to tackle the approximation error on the value distribution, a recent line of work has focused on learning the optimal mechanism assuming the seller has access to a batch of i.i.d. examples of bidders' valuations \citep{MorgenRough2016,paes2016field}.  

Several works have already shown that if bids are used in order to design the mechanism, the bidders should no longer bid truthfully \citep{tang2016manipulate,kanoria2017dynamic,mohri2015revenue,abeille2018explicit}. However, they do not exhibit  strategies which work in the general setting with arbitrary value distributions  and do not address cases where bidders only have partial and noisy access to the competition's distribution. Indeed, in practice, ad platforms disclose little information to participants.

Our paper introduces a simple, general and robust method to design bidding strategies adaptive to prior-dependent revenue maximizing auctions. This method works with very general value distributions, asymmetries between bidders and various revenue maximizing mechanisms. It also tackles practical settings where no prior information on the competition is available. It should enable practitioners to easily improve their bidding strategies.

\subsection{Framework and related work}

Starting with the seminal work of \cite{Myerson81}, a rich line of works indicates the type of auctions that is revenue-maximizing for the seller. In the case of symmetric bidders \citep{Myerson81}, one revenue maximizing auction is a second price auction with a reserve price equal to the monopoly price, i.e, the price $r$ that maximizes $r(1-F(r))$ ($F$ being the value distribution of the bidder). However, in most applications, the symmetry assumption is not satisfied \citep{Golrezaei2017}. In the asymmetric case, the Myerson auction is optimal but difficult to implement in practice \citep{morgenstern2015pseudo}. In this case, a second price auction with well-chosen reserve prices guarantees at least $1/2$ of the optimal revenue \citep{Hartline2009}. 
 
Revenue maximizing auctions play a big role in the internet economy \citep{medina2014learning,paes2016field} because there exist a lot of heterogeneities between bidders and a low number of participants per auction \citep{mohri2015revenue}. In this context, they have a clear impact on sellers' revenues since the ratio between revenues of the Myerson auction (the optimal revenue-maximizing auction) and those of a second price auction without a reserve price (the welfare-maximizing auction) is of order $1 - 1/n$ with $n$ the number of bidders \citep{fu2015randomization}.

Practical implementation of revenue-maximizing mechanisms requires that the seller knows the bidders' value distributions beforehand, which practically may not be the case. More precisely, assume the valuation of a bidder $v_i \in \mathbb{R}$ is drawn from a specific distribution $F_i$;  a bidding strategy for bidder $i$ is a mapping $\beta_i$ from $\mathbb{R}$ into  $\mathbb{R}$ that indicates  the actual bid $B_i=\beta_i(v_i)$ when  the value is $v_i$. As a consequence, the distribution of bids $F_{B_i}$ is  the push-forward of $F_i$ by $\beta_i$. We can highlight different classes of auction problems depending on the information available to the seller that she might use to optimize her mechanism. 
\begin{enumerate}
\item[(S1)] $F_{i}$ is known to the seller. This is the traditional setting studied in Myerson's seminal paper where he assumed that $\{F_{i}\}$'s are common knowledge among the bidders and the seller.
\item[(S2)] $F_{B_i}$ is known to the seller. It is an idealized setting where the seller has no uncertainty about $F_{B_i}$, e.g. because she receives an infinity of i.i.d. bids from the buyers. 
\item[(S3)] the seller has only access to a finite number of examples of bids. She can only compute an approximation of $F_{B_i}$ denoted by $\hat{F}_{B_i}$. This is the most realistic case.
\end{enumerate}
The last setting (S3) best describes real-world practice and the objective of most of mechanisms is to make sure that bidders are truthful. Indeed, the bid distribution is then equal to the value distribution and the seller is back to the full information setting (S1) required to implement revenue-maximizing mechanisms. Hence, designing incentive compatible mechanisms is a crucial requirement to elicit the bidders' private value distributions. The remaining question of empirical estimation was theoretically addressed by \cite{cole2014sample,huang2018making,devanur2016sample} looking at the sample complexity of a large class of auctions assuming access to i.i.d. examples of the value distribution.

In order to design such incentive compatible mechanisms, sellers have been relying on different assumptions. In the traditional setting of auction theory, sellers assume that bidders are myopic and do not optimize their long term-utility \citep{OstSch11,paes2016field}. However, in the context of  internet auctions, given the volume and frequency of auctions - billions a day -, sellers cannot assume myopic behavior by the bidders.
In this context, one needs to account for the dependency introduced by the seller on the bidder strategy when using past bids to adapt the mechanism (for instance by optimizing the reserve price). In such cases, non-myopic bidders optimizing their \textit{long-term expected utility} have an incentive to be strategic against this adaptation of the mechanism. 
To solve this issue, \cite{amin2014repeated,mohri2015revenue,golrezaei2018dynamic} exhibited mechanisms that are incentive compatible (up to a small number of bids) under the assumption that bidders are \emph{almost} myopic or impatient -- i.e. they have a fixed discount on future utilities. Unfortunately, it comes at the cost of introducing an asymmetry between the bidders having a discounted long-term utility, and the seller having an undiscounted long-term revenue (in effect being infinitely patient). Another way to prevent bidders from being strategic is to adapt the mechanism (e.g. reserve price) based on the competition of a bidder rather than based on the bidder themselves \citep{ashlagi2016sequential,kanoria2017dynamic,epasto2018incentive}. A limitation of this type of approach is the need to rely on a (partial) symmetry of the bidders: any bidder needs to have competitors with (almost) the same value distribution as her. In particular and as noticed in \cite{epasto2018incentive}, it cannot handle the existence of any dominant buyer, i.e., a buyer with higher values than the other bidders. This is a limited setting as revenue-optimizing mechanisms are mostly needed when the buyers are heterogenous, for otherwise the competition itself contributes to optimizing the revenue of the seller \citep{bulow1996auction}. In the real-world setting of online advertising, with asymmetric bidders and no specific asymmetry between seller and buyers on future utilities, none of these mechanisms ends up being able to enforce truthful bidding. This is illustrated by recent works proposing a method to empirically detect when a mechanism is not truthful from the point of view of a non-myopic bidder \citep{lahaie2017testing,feng2019online,DengLa19}. In this setting, this leaves us with an important question: 
\begin{center}
What should ``good" bidders' strategies be?
\end{center}
Closely related to our work, \citet{tang2016manipulate} derive a Nash equilibrium in our specific setting but their approach does not enable them to derive equilibrium in some restricted class of strategies that are widely used in practice and do not tackle any robustness issues related to their strategies. The main objective of our work is to exhibit and prove the existence of simple and robust strategies that can be used by practitioners facing a smart data-driven selling mechanism. Understanding possible simple strategic behaviours will also help sellers to design robust strategy-proof selling mechanisms. On a technical level, our approach is based on calculus of variations ideas and appears to be new in the auction context. 
\subsection{Main contributions}

We introduce a method to derive strategies for the bidder in the context of a seller using her past bids to adapt the mechanism towards a revenue-optimization objective. We mainly focus in this paper on the case where the mechanisms are lazy second price auctions with monopoly reserve price and extend some of the results to other classes of auctions such as the Myerson auction, the eager second price auction with monopoly price or the boosted second-price auction.
\iffalse  More formally, we consider a two-stage game where the seller is switching to a revenue-maximizing objective after gathering examples of bids during the first stage \citep{amin2014repeated,lahaie2017testing}.  The seller is optimizing the reserve prices according to  $F_{B_i}$  or $\hat{F}_{B_i}$ that she observed during the first stage. We focus on the specific case where the bidders are only optimizing their utility in the second stage. We assume that bidders are risk-neutral and seek to maximize their utility based on the information available to them: 
\begin{enumerate}
\item[(B1)] either they know the class of mechanisms they are facing and only their own distribution $F_{i}$. They do not have any side information. This is the realistic problem corresponding to internet auctions since in practice ad platforms disclose little information.

\item[(B2)] or  they know the class of mechanisms they are facing, their own value distribution and the bid distribution of the competition (i.e. the distribution of the maximum bid of their competitors.)
\end{enumerate}
Based on this classification of auction problems, we introduce the following practical objective~: bidders aim at maximizing in the ``long run''  their own expected utility given that they know the class of mechanisms used by the auctioneer, their own value distribution and that their bids might be used to optimize the mechanism.
\fi
\subsubsection{Timing of the game}
We first consider an idealized game where the buyer discloses their bid distribution and the seller optimizes their auction mechanism based on this distribution. For concreteness, consider the case of second price auctions. In the classical setting of auction theory, the buyer is asked to reveal their bid distribution first; facing ``truthful auctions'', they reveal their value distribution. The seller then optimizes their mechanism based on this information, finding an optimal reserve price for this buyer. This is a Stackelberg game, as the two players do not play at the same time. In this instance, the seller is the leader and the buyer is the follower. Most of the literature on optimal auctions is focused on this version of the Stackelberg game. 

However, if the buyer knows that the seller is going to find an \textbf{optimal} mechanism, and hence will optimize the auction based on the information given by the buyer's bid distribution, the buyer can anticipate this optimization to increase their utility. The order of the Stackelberg game is then reversed. The buyer becomes the leader and the seller the follower. The buyer reveals their bid distribution knowing the optimization problem that the seller will solve. In second price auctions with reserve prices, the buyer has an incentive to disclose a bid distribution that may be different from their value distribution as they then will be facing a more favorable reserve price by doing so. Our paper is focused on this version of the Stackelberg game, as we focus on the bidder standpoint. 

More formally, the timing of the game we consider is the following :
\begin{enumerate}
\item the seller chooses a mapping $\mathcal{M}$ from $F_{B_i}$ to a specific auction mechanism $\mathcal{M}(F_{B_i})$,
\item based on this mapping, each buyer chooses a bidding strategy $\beta_i$,
\item the payoff of the buyers is their utility under $(F_i, \mathcal{M}(F_{B_i}), \beta_i)$ and the seller's payoff is the revenue under this mechanism.
\end{enumerate}

This objective is particularly relevant in modern applications as most of the data-driven selling mechanisms are using large batches of bids as examples to update their mechanism. In practice however, the problem is slightly different. Classically, the seller could first sell goods in a second price auction without reserve price \cite{amin2014repeated,mohri2015revenue,golrezaei2018dynamic}. The buyer would then be incentivized to reveal their value distribution - through participating in a large number of those auctions. And the seller would then optimize the reserve price based on this value distribution for the remaining auctions. This is again a Stackelberg game where the seller is the leader. However, if the buyer is aware that their bid distribution in the first phase is going to be used in a second phase to run seller-optimal auctions, the buyer can take the lead in the Stackelberg game by anticipating the optimization performed by the seller. 
\iffalse
More formally, the seller uses the distributions of bids $F_{B_i}$ to choose a specific auction mechanism $\mathcal{M}(F_{B_i})$  among a given class of mechanisms $\mathcal{M}$. The objective of \textit{a long-term strategic bidder} is to find her strategy $\beta_i$ that maximizes her expected utility when $v_i \sim F_i$, she bids $\beta_i(v_i)$ and the induced mechanism is $\mathcal{M}(F_{B_i})$. 
In terms of game theory, these interactions are a game between the seller - whose strategy is to pick a mechanism design that  maps bid distributions to reserve prices - and the bidders - who chose bidding strategies. This game can be interpreted as a Stackelberg game where bidders are the leaders since they act first. Our overarching objective is to derive the best-response (or at least provably improved response), for a given bidder $i$,  to  the strategy of the seller (i.e. a given mechanism) and to the strategies of the other bidders (i.e. their bid distributions) up to available knowledge on the competition. 
\fi
\begin{table}[]
\centering
\begin{tabular}{c|c|c|c|c|c|c|c|c}  
& \multicolumn{4}{c|}{Optimal reserve price} & \multicolumn{4}{c}{Utility} \\ \cline{2-9}
&K=1 &K=2&K=3&K=4  &K=1 &K=2&K=3&K=4\\ \hline
\multirow{2}{*}{Truthful bidding} & \multirow{2}{*}{0.5}& \multirow{2}{*}{0.5}& \multirow{2}{*}{0.5}&  \multirow{2}{*}{0.5}& \multirow{2}{*}{1/8} & \multirow{2}{*}{1/12} & \multirow{2}{*}{11/192}& \multirow{2}{*}{13/320} \\ 
 & &  &  & &&&& \\ \hline
\multirow{2}{*}{Zero bidding} &\multirow{2}{*}{0.0}&\multirow{2}{*}{0.0}&\multirow{2}{*}{0.0}& \multirow{2}{*}{0.0}  & 1/2 & 0.0 &  0.0 & 0.0\\ 
 &&&&  & (+400\%)  & (-100\%)  & (-100\%) & (-100\%) \\ \hline
 \multirow{2}{*}{Divide values by 2} &\multirow{2}{*}{0.25}&\multirow{2}{*}{0.25}&\multirow{2}{*}{0.25}& \multirow{2}{*}{0.25}  & 1/4 & $\approx 0.094$ &  $\approx 0.036$& $\approx 0.015$\\ 
 &&&&  & (+100\%)  & (+13\%)  & (-37\%) & (-63\%) \\ \hline
Thresholded at & \multirow{3}{*}{0.25}& \multirow{3}{*}{0.25}& \multirow{3}{*}{0.25}& \multirow{3}{*}{0.25}  & \multirow{2}{*}{1/4} &  \multirow{2}{*}{$\approx 0.132$}  & \multirow{2}{*}{$ \approx 0.076$}  & \multirow{2}{*}{$\approx 0.048$}\\
the monopoly price & &&&&&&&\\
(Theorem \ref{thm:settingReserveValToZeroImprovesPerformance}) &  &&&& (+100\%)  & (+57\%)  & (+33\%) & (+20\%) \\ \hlineB{4.0}
Optimal  regularity-& \multirow{2}{*}{0.0}& \multirow{2}{*}{0.162}& \multirow{2}{*}{0.204}& \multirow{2}{*}{0.22}  & \multirow{2}{*}{1/2} & \multirow{2}{*}{$\approx 0.147$}  & \multirow{2}{*}{$\approx 0.079$}  & \multirow{2}{*}{$\approx 0.049$}\\
preserving strategies&&&&  &  & &  & \\ 
(Theorem \ref{thm:quantificationOptimalThresholding})&&&&  & (+400\%)  & (+76\%)  & (+38\%) & (+21\%) \\ 
\hline
\end{tabular}
\caption{Comparison of the utility of the strategic bidder between the truthful strategy, the strategy corresponding to bidding zero for any values, the linear strategy dividing values by two, the strategy introduced in Theorem \ref{thm:settingReserveValToZeroImprovesPerformance} and the optimal regularity-preserving strategies for each number of competitors (derived from Theorem \ref{thm:quantificationOptimalThresholding}). The first four strategies are fixed and do not require knowledge of the competition to be computed. The last one is competition-specific and exact knowledge of the distribution followed by the highest bid of the competition is needed to compute it. For this example, bidders' value distributions are $\mathfrak{U}[0,1]$ and  opponents are assumed to bid truthfully.}
\label{table_intro}
\end{table}
\subsubsection{A simple textbook example}
To introduce our contribution, we first consider the posted price setting where $K=1$ bidder plays against one seller. We assume, for simplicity of this introductory example, that bidder's value distributions $F_{X_i}$ is $\mathfrak{U}[0,1]$, i.e. uniform on the interval [0,1]. Let initially consider that the bidder is bidding truthfully, i.e, $\beta_i = Id$. In this case, $F_{B_i} = F_{X_i}$ and the seller will set as reserve price the monopoly price by maximizing the monopoly revenue $r(1-F_{X_i}(r))$. This monopoly price is equal to 0.5 in the case of $\mathfrak{U}[0,1]$.  Note that this maximization problem is computationally simple as the monopoly revenue is a concave function if the value distribution is regular. The bidder can obviously do better. If he bids all the time zero  (or $\varepsilon$ arbitrarily close to zero), $F_{B_i}$ will be equal to a point mass at zero. Through computing the optimal reserve price corresponding to $F_{B_i}$, the seller chooses zero, obviously maximizing bidder's utility. The problem we consider derives from a simple extension of this example to the case of K bidders. In a lazy second price auction, the optimal reserve price for each bidder is still the monopoly price. Yet, as soon as there is some competition, bidders can not bid zero as they get zero utility in this case.  They have to tradeoff between beating the competition and decreasing their reserve price. 

Our first result consists in deriving a simple strategy which guarantees to the bidder an increase in utility compared to the truthful strategy for any distributions of the competition. This increase depends on the distribution of the competition. Yet, by playing this strategy, the bidder is sure to do better than by bidding truthfully. This is an important practical result as in many ad platforms, bidders have to bid without knowing the distribution of the competition. This strategy, that we call \textit{thresholding at the monopoly price}, has also the key property of making simple the optimization problem of the seller, i.e. if $F_{X_i}$ is regular, $F_{B_i}$ induced by
this strategy on  $F_{X_i}$ is also regular. We say that this strategy is regularity-preserving.

\begin{definition}[Regularity-preserving strategy] Consider a bidder with a regular value distribution $F_{X_i}$. We say that a bidding strategy $\beta_i$ is regularity-preserving, if the bid distribution $F_{B_i}$ induced by $\beta_i$ on $F_{X_i}$ is a regular distribution.
\label{def:RP}
\end{definition}

As shown in Table \ref{table_intro}, we then extend our first result by deriving, for any given regular distribution, the optimal strategy in the class of increasing and regularity-preserving strategies. To compute this strategy, the bidder has to know exactly the distribution of the competition. Our result enables us to compute analytically by how much the bidder increases his utility when knowing precisely the distribution of the competition.

\subsubsection{Our main results}

The paper is organized as follows. Section \ref{sec:improvingAnyBiddingStrategy} provides a walkthrough of the main technical ideas supporting our work to provably improve the response of one bidder independently from the behavior and knowledge of other bidders. In Theorem 1, we define this new strategy, independent of the distribution of the competition, which enables to increase the strategic bidder's utility compared to the truthful strategy for any distribution of the competition. This theoretical study introduces a practical method to design bidding strategies in revenue-maximizing auctions that we call \textit{thresholding the virtual value}. In Section \ref{sec:best_response_with_competition_knowledge}, we generalize this result by exhibiting, for a given distribution of the competion, what is the optimal strategy in the class of regularity-preserving strategies (Theorem 2). We show that this strategy belongs to the class of strategies that we introduced in Section \ref{sec:improvingAnyBiddingStrategy}. This optimal strategy depends on the distribution of the competition and corresponds to the optimal tradeoff between decreasing the reserve price and beating the competition. 

In Section \ref{sec:bayes_nash}, we show that our results can be made robust to other strategic bidders and in Section \ref{sec:empirical_estimation} to  sample-approximation error by the seller, tackling the case where only an estimator $\hat{F}_B$ is available to the seller.  If the seller uses a data-driven approach as proposed in \cite{paes2016field} to set the reserve prices in a lazy second-price auction, we show under what conditions thresholding bidding strategies can improve buyer’s utility.

\section{Improving over the truthful strategy for any distributions of the competition}
%\section{Improving any bidding strategy with respect to reserve prices}
\label{sec:improvingAnyBiddingStrategy}
In this section, we show how to improve the strategy of one bidder when he does not have any information about the competition. We assume that the strategies of the other bidders are fixed and do not depend on the strategic bidder's strategy and that the seller has perfect knowledge of the bid distributions to choose the mechanism. We relax this latter assumption in Section \ref{sec:empirical_estimation}. 

\subsection{Notations and setting}
We recall that $F_{i}$ is the value distribution of bidder $i$ and $\beta_i : \mathbb{R} \to \mathbb{R}$ her strategy that maps values to bids. The corresponding distribution of bids is then $F_{B_i} = \beta_i \sharp F_i$, the push-forward of $F_i$ w.r.t.\ $\beta_i$.  Notice that we have implicitly identified the distribution $F_i$ (resp.\ $F_{B_i}$) with its cumulative distribution function (cdf) and use both terms exchangeably. We use $f_i$ (resp.\ $f_{B_i}$) for the corresponding probability density function (pdf). 

We denote by $\psi_{B_i}$ the corresponding virtual value function defined as
\begin{equation*}
\psi_{B_i}(b) = b - \frac{1-F_{B_i}(b)}{f_{B_i}(b)}\;.
\end{equation*}
We use $G_i$ (resp. $G$ when there is no ambiguity) for the distribution of the highest bid of the competition to bidder $i$. We call $g_i$ (resp. $g$) the associated density.

Throughout the paper we assume that the random variables associated with $F_i$ and $G_i$ are drawn independently. 

We denote by $\mathfrak{U}$ the uniform distribution and $\mathfrak{U}[0,1]$ the uniform distribution on [0,1]. 

For each theorem, we provide quite minimal assumptions needed on the value distribution for the proof to go through. However, as a summary, all the main results on the best response of the strategic bidder are true provided 
\begin{itemize}
\item \textbf{A1}~: the value distribution has $1+\eps$ moments, $\eps>0$ 
\item \textbf{A2}~: the value distribution $F$ has a density $f$, with $f>0$ on the support of $F$ (in particular $F$ has no atoms and its cumulative distribution function is continuous)  
\item \textbf{A3}~: the virtual value is increasing and crosses 0 (though in many instances we really only need that it crosses 0 exactly once - in which case we denote the assumption by \textbf{A3w}) 
\end{itemize}
This includes among others distributions the uniform and the exponential distributions. Nevertheless, for most of theorems, we do not need any regularities assumptions on the value distribution. Our theorem on the Bayes-Nash equilibrium (Theorem \ref{thm:revBuyerInThreshStrategy}) further requires (\textbf{A4}) a compact support for the distribution as well as continuity of the density at the edges of the support. We however explain after the theorem how to extend it to Generalized Pareto and log-normal distributions (when the latter distributions are regular \cite{ewerhart2013regular}). Finally, our theorem on robustness to approximation error of the seller (Theorem \ref{thm:epsAndMonopolyPriceByERM}) requires (\textbf{A5}) in its stated form that the density be bounded below under the monopoly price. In conclusion, all our results apply for the most common distributions used in auction theory, provided they are truncated in an infinitesimal neighborhood near 0 and away from infinity (if needed) to avoid minor technical problems in the proofs. This does not compromise their wide applicability.

We assume that the seller runs a lazy second price auction with monopoly reserve price computed according to the distribution of bids she is observing or knows. We recall that in a ``lazy''  second price auction, the item is attributed to the highest bidder, if she clears her reserve price, and not attributed otherwise; the winner pays the maximum of the second highest bid and her reserve price.  
\iffalse
It is known that the optimal reserve price of bidder $i$ is  the monopoly price equal to  $\argmax_r r(1-F_{B_i}(r))$, or equivalently\footnote{at least for regular distributions, i.e., when $\psi$ is increasing} to $\psi_{B_i}^{-1}(0)$. 
As a consequence, it is natural \nek{why?} to assume that the strategy of bidder $i$ does not impact the strategy of other bidders (that can be either myopic or not) and from now on, we assume that bids are independent.
\fi

\subsection{Improving any strategies for any competition distributions}
In the initial example presented in the introduction, we considered several classical shading strategies, e.g. zero bidding or linear shading, that can be used by a strategic bidder. As shown in Table \ref{table_intro}, they are suboptimal and perhaps more importantly do not guarantee a positive uplift relative to the truthful strategy for any distribution of the competition. For instance, in the textbook example with $\mathfrak{U}[0,1]$, dividing bids by two gets a positive uplift with two bidders but not with three. This is a severe limitation since in some situations, the seller prevents bidders from computing the distribution of the competition by e.g. not saying whether the bidder paid the second highest bid or their reserve price. In practice, approximating the distribution of the competition also requires an important infrastructure. We show that there exists a simple strategy that \emph{guarantees a positive uplift in utility compared to the truthful strategy for any distributions of the competition}. This strategy is of practical interest since bidders only need to know their own value distribution to compute this strategy. We will extend this result in Section \ref{sec:best_response_with_competition_knowledge} by showing that the best response in this setting is a simple extension of the strategy that we now introduce.

When the reserve price is computed from $F_{B_i}$ -  the bid distribution induced by using $\beta$ on $F_{X_i}$ -  we have to make a distinction between the reserve price $r_\beta$ and the reserve value $x_\beta$.

\begin{definition}[Reserve value]
Consider a non-decreasing strategy $\beta$. We call reserve value $x_\beta$  the smallest value above which the seller accepts bids. 
\end{definition}
If the bidder bids truthfully, their reserve value is equal to their reserve price. If $\beta$ is increasing and $r_\beta$ is the reserve price associated with the strategy $\beta$, we have $x_\beta = \beta^{-1}(r_\beta)$. If $F_X = \mathfrak{U}[0,1]$, if we consider $\beta(x) = x/2$, we have $r_\beta = 0.25$ and $x_{\beta} = 0.5$. By dividing bids by two, the strategic bidder decreases their reserve price but does not change the reserve value~: it is the same as if they were  bidding truthfully.

By contrast, we now show that for any distributions of the competition, the bidder can improve a specific bidding strategy by giving an incentive to the seller to accept all bids. It does not decrease the payment of the strategic bidder and increases his utility for any distributions of the competition. % This theorem does not require any assumptions on the value distribution, beside its existence (implied by Assumption \textbf{A2}).

\begin{theorem}\label{thm:settingReserveValToZeroImprovesPerformance}
Let $\beta_r$ be an increasing strategy with associated reserve value $r>0$ in a lazy second price auction. Suppose Assumption \textbf{A2} applies to $F$ and that the left-end point of its support is 0. Suppose the bid distribution associated with $\beta_r$ has a virtual value. Assume that the other bidders' strategies are fixed. Then there exists another bidding strategy $\tilde{\beta}_r$ such that for any distributions of the highest bid of the competition: 
\begin{enumerate}
\item A reserve value associated with $\tilde{\beta}_r$ is 0. $\tilde{\beta}_r$ is increasing.
\item $\mathcal{U}_i(\tilde{\beta}_r)\geq \mathcal{U}_i(\beta_r)$, i.e., the utility of bidder $i$ is higher;
\item $P_i(\tilde{\beta}_r)\geq P_i(\beta_r)$, i.e., the payment of bidder $i$ to the seller is also higher,
\end{enumerate}
The following continuous function fulfills these conditions:
\begin{equation}
\tilde{\beta}_r(x)=\left(\frac{\beta_r(r)(1-F_{X_i}(r))}{1-F_{X_i}(x)}\right)\mathds{1}\{x<r\}+\beta_r(x)\mathds{1}\{x\geq r\} \label{eq:DefNewThreshBidFunction}
\end{equation}
\end{theorem}
Having reserve value equal to zero means that the seller accepts all bids of the strategic bidder. It also means that the reserve price is equal to the minimum bid of the strategic bidder. This result can be applied to improve any preexisting shading strategy. A most important case is to apply this theorem to the truthful strategy, showing that there exists a strategy improving the truthful strategy regardless of the competition distribution. The proof is in Appendix. We now explain why we can improve any strategy in this setting without knowing the distribution of the competition.  Myerson's Lemma is a key element in this understanding.

\subsubsection{The fundamental Myerson lemma}
\iffalse
\begin{figure}[h!]
\begin{tabular}{ccc}
\includegraphics[width=.33\linewidth]{r025}&
\includegraphics[width=.33\linewidth]{r075}&
\includegraphics[width=.33\linewidth]{R05}
\end{tabular}
\caption{\textbf{Change in bidder payment as a function of the reserve price.} The value distribution of the bidder is uniform [0,1], for the sake of illustration. XHer virtual value is therefore equal to $\psi(x) = 2x -1$, and is represented by the blue line. The dashed red vertical line corresponds to the current reserve price. We picked $G=1$, i.e. no competition, for the sake of clarity of the plot. The bidder payment is equal to the area under the curve. The seller has no incentive to set the reserve price lower than $\psi_i^{-1}(0)$ since it decreases her total expected payment (because of the negative contribution of the red area). It is also clear that $r \ge \psi_i^{-1}(0)$ is suboptimal since it results in lost revenue for the seller.}
\label{fig:visualOptimalR} % I can do without the label too
\end{figure}
\fi
Myerson's lemma  \citep{Myerson81} is a fundamental result in auction theory.
\begin{lemma}[Integrated version of the Myerson lemma]\label{Myerson_lemma}
Assume that  $F_{B_i}$ has a density (which is positive everywhere) and finite mean, i.e. Assumptions \textbf{A1} (with $\eps=0$) and \textbf{A2} are satisfied. Suppose that bidder $i$'s bids are independent of the bids of the other bidders and denote by $G_i$ the cdf of the maximum of the other bidders' bids. Suppose a lazy second price auction with reserve price for bidder i denoted by $r$ is run. Then the payment $M_i$ of bidder $i$ to the seller can be expressed as 
\begin{equation*}
M_i(\beta_i) = \mathbb{E}_{B_i \sim F_{B_i}}\bigg(\psi_{B_i}(B_i)G_i(B_i)\textbf{1}(B_i \geq r)\bigg)\;.
\end{equation*}
If the other bidders are bidding truthfully, $G_i$ is the distribution of the maximum value of the other bidders.
\end{lemma}
The formal proof is in Appendix. Lemma \ref{Myerson_lemma} yields that it is enough to consider the virtual value to deduce the payment in the lazy second price auction. This explains why the optimal reserve price is equal to $\psi_{B_i}^{-1}(0)$ when $\psi_{B_i}$ crosses 0 exactly once and is positive beyond that crossing point \textbf{A3w} (that is of course the case when $\psi_{B_i}$ is increasing and crosses 0 \textbf{A3})~: the derivative of $M_i(\beta_i)$ with respect to the reserve price $r$ has the opposite sign as that of $\psi_{B_i}(r)$.

% The bidder payment is equal to the area under the curve (we removed the dependence on G for the sake of clarity). The seller has no incentive to set the reserve price lower than $\psi_i^{-1}(0)$ since it decreases her total expected payment (because of the negative contribution of the red area). It is also clear that $r \ge \psi_i^{-1}(0)$ is suboptimal since it results in lost  revenue for the seller.

Using the notations introduced previously, it is optimal for the seller to choose as reserve price for bidder $i$ the monopoly price corresponding to her bid distribution, and Lemma \ref{Myerson_lemma} implies that the expected payment of bidder $i$ in the optimized lazy second price auction is equal to 
\begin{equation*}
M_i(\beta_i) = \mathbb{E}_{B\sim F_{B_i}}\bigg(\psi_{F_{B_i}}(B)G_i(B)\textbf{1}(B \geq \psi_{B_i}^{-1}(0))\bigg)\;.
\end{equation*}
In order to simplify the computation of the expectation and remove the dependence on $B_i$, we rewrite this expected payment in the space of values using the fact that the strategic bidder is using an increasing strategy $\beta_i$. We will only consider increasing strategies in the rest of the paper. To do so, we define:
\begin{equation*}
h_{\beta_i}(x) \triangleq \psi_{F_{B_i}}(\beta_i(x))  
\end{equation*}
With this new notation, we can rewrite the expected payment of the strategic bidder $i$
\begin{equation*}
M_i(\beta_i) = \mathbb{E}_{X_i \sim F_{i}}\bigg(h_{\beta_i}(X_i)G_i(\beta_i(X_i))\textbf{1}(X_i \geq x_\beta)\bigg)\;.
\end{equation*}
and derive her expected utility as a function of $\beta_i$ as
\begin{equation}\label{equ:utility}
U(\beta_i) = \mathbb{E}_{X_i \sim F_{i}}\bigg((X_i-h_{\beta_i}(X_i))G_i(\beta_i(X_i))\textbf{1}(X_i \geq x_\beta)\bigg)\;.
\end{equation}
where $x_\beta$ is \emph{the reserve value}. If $h_{\beta_i}$ crosses 0 exactly once and is positive beyond that crossing point, i.e. it satisfies \textbf{A3w},  $x_\beta =  h_{\beta_i}^{-1}(0)$.  If we call $r_i=\psi_{F_{B_i}}^{-1}(0)$ the \emph{reserve price} of bidder $i$ and $\beta_i$  increasing , the \emph{reserve value} is equal to $\beta_i^{-1}(r_i)$. 

If we consider only increasing differentiable strategies, and we denote by  $\mathcal{I}$ the class of  such functions, the problem of the strategic bidder is therefore to solve  $ \sup_{\beta \in \mathcal{I}} U(\beta) $ with $U$ defined in Equation \eqref{equ:utility}. This equation is crucial, as it indicates that optimizing over bidding strategies can be reduced to finding a distribution with a well-specified  $h_\beta(\cdot)$. Our results extend to the case where the strategies are increasing and differentiable except at finitely many points, as we only need $bF_B(b)$ to be absolutely continuous for the previous result to go through.

A crucial difference between the long term vision and the classical, myopic (or one-shot)  auction theory is that in our setup bidders maximize expected utility globally over the full support of the value distribution. In the classical myopic setting, bidders determine their bids so as to maximize their expected utility at each value. In our setup, the strategic bidder also accounts for the computation of the reserve price, a function of her global bid distribution. He might therefore be willing to sometimes over-bid (incurring a negative utility at some specific auctions/values) or underbid (lose some auctions that he would have won otherwise) if this reduces her reserve price. Indeed, having a lower reserve price increases the utility of other auctions. Lose small to win big. In other words, the strategy trades-off ex-post individual rationality (IR) for higher utility (of course ex-ante IR still holds). This reasoning makes sense only with multiple interactions between bidders and seller.

\subsubsection{Thresholding the virtual value}
\label{subsec:thresholding_virtual_value}
\begin{figure}[t]
\centering
\begin{tabular}{cc}
\includegraphics[width=.40\linewidth]{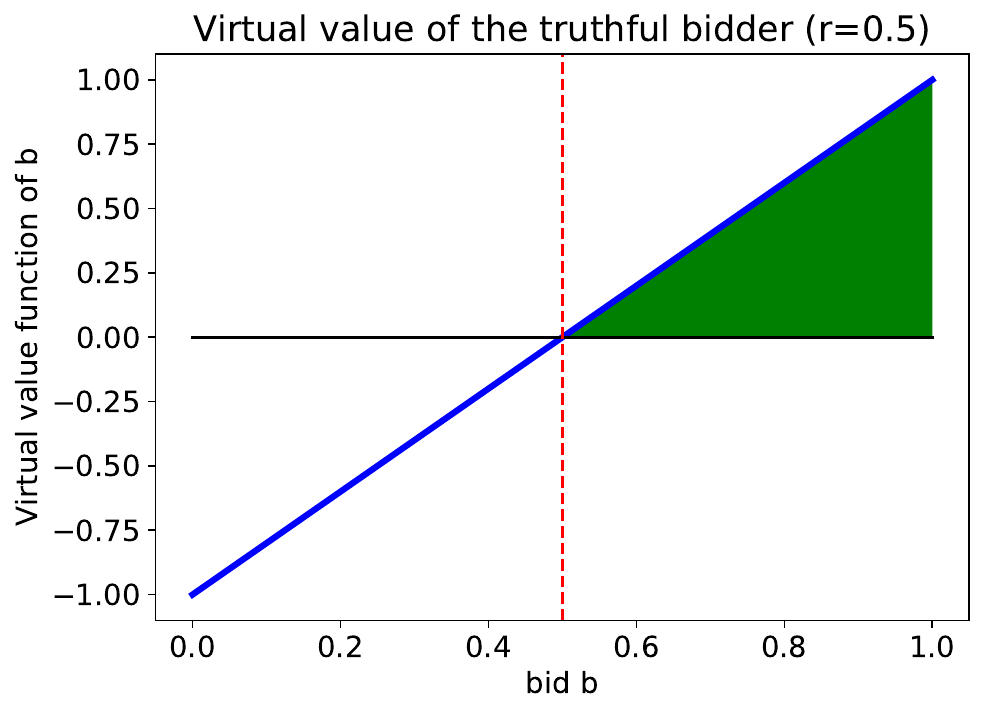} &
\includegraphics[width=.40\linewidth]{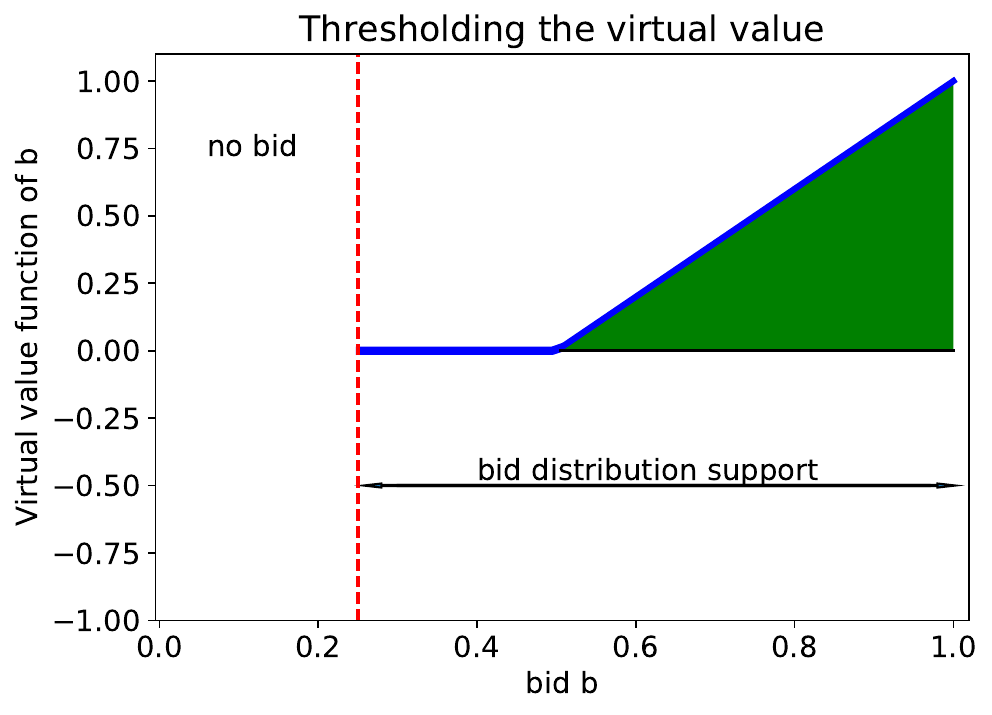}
\end{tabular}
\caption{\textbf{Virtual value of truthful bidder vs. strategic bidder.} The value distribution of the bidder is $\mathfrak{U}[0,1]$, the standard textbook example used for the sake of illustration. Her virtual value is therefore equal to $\psi(x) = 2x -1$, and is represented by the blue line. The dashed red vertical line corresponds to the current reserve price. The green area corresponds to the bidder's payment if we picked $G=1$, i.e. no competition, for the sake of clarity of the plot. The left-hand side corresponds to truthful bidding, the right-hand side to strategic behavior. In both cases, the blue line corresponds to $\psi_{B}$.}
\label{fig:fig1vValueTruthvsStrat} % I can do without the label too
\end{figure}

To explain why we can easily improve a bidding strategy, let us describe the following elementary example (see also Figures \ref{fig:fig1vValueTruthvsStrat} for graphical illustrations). We assume that the value distribution of the bidder is uniform between 0 and 1 (denoted $\mathfrak{U}[0,1]$, a standard example in \cite{krishna2009auction}, see e.g. Example 2.1 and subsequent chapters). If the strategic bidder were bidding truthfully, her virtual value would be negative below 0.5 and the seller would set the reserve price to 0.5. What would happen if the bidder were able to send a virtual value equal to zero below 0.5? Then the seller would not have any incentives to block some of the bids since the virtual value is non-negative everywhere. Notice that, since the virtual value is zero below 0.5, the seller receives exactly the same expected payment as in the previous setup.

We define \emph{a welfare-benevolent seller} as a seller who  - when indifferent from a revenue standpoint between two reserve prices -  chooses the lowest one, i.e, the one yielding the maximum welfare. This is a standard concept in game theory. In the framework where the seller has a cost to sell the item \cite{balseiro2017dynamic}, it corresponds to a cost-to-sell of zero.

If the seller is not benevolent, instead of looking for a strategy such that $\psi_B(x) = 0$ on $[0,0.5]$, the buyer will try to satisfy $\psi_B(\beta(x)) = \epsilon$ with $\epsilon$ small. In that case the seller has a strict incentive to take all bids and hence lower the reserve price. 

We call this technique \textbf{thresholding the virtual value}: finding a bidding strategy such that the virtual value of the induced distribution is equal to zero (or to a threshold $\epsilon$, for arbitrary small $\epsilon>0$ if the seller is not welfare benevolent) below the current reserve price.

We now show formally how to find a bidding strategy such that the virtual value of the induced bid distribution is equal to zero below a certain threshold.
\subsection{Deriving formally the corresponding bidding strategy}
Before carrying on with reasoning on the virtual value, such as in our motivating example, we need to ensure we can find the corresponding strategy $\beta_i$ that will expose a bid distribution $F_{B_i}$ with the corresponding virtual value to the seller. The two following technical lemmas show how to deduce $\beta_i$ from a given $h_{\beta_i}$.
%To derive the corresponding bidding strategy, we present two important technical lemmas. 
\begin{lemma}\label{definition_psi}
Suppose $B_i=\beta_i(X_i)$, where $\beta_i$ is increasing and differentiable and $X_i$ is a random variable with cdf $F_i$ and pdf $f_i$ satisfying \textbf{A2}. Then 
\begin{equation}\label{eq:ODEPhiG}
h_{\beta_i}(x_i)\triangleq \beta_i(x)-\beta_i'(x)\frac{1-F_{i}(x)}{f_{i}(x)} = \psi_{F_{B_i}}(\beta_i(x)) \;.
\end{equation}
\end{lemma}
\begin{proof}
We have $\psi_{F_{B_i}}(b) = b - \frac{1 - F_{B_i}(b)}{f_{B_i}(b)}$ with $F_{B_i}(b) = F_{i}(\beta_i^{-1}(b))$ and $f_{B_i}(b) = f_{i}(\beta_i^{-1}(b)/\beta_i'(\beta_i^{-1}(b))$.
Then, 
$h_{\beta_i}(x) = \psi_{B_i}(\beta_i(x)) = \beta_i(x) - \beta_i'(x)\frac{1-F_{i}(x)}{f_{i}(x)}\;.$
\end{proof}
The above results holds when $\beta$ is increasing, continuous, and differentiable except at finitely many points. $h_{\beta_i}$ in Equation \eqref{eq:ODEPhiG} is then the definition at all the points of differentiability of $\beta_i$. Note that Lemma \ref{Myerson_lemma} then applies as it relies on integration by parts which is valid for functions that are continuous and differentiable except at finitely many points.

The second lemma shows that for any function $g$ we can find a function $\beta$ such that $h_\beta = g$.

\begin{lemma}\label{lemma:keyODEs}
Let $X$ be a random variable with cdf $F$ and pdf $f$ satisfying \textbf{A2}. Let $x_0$ be in the support of $X$, $C\in\mathbb{R}$ and $g:\mathbb{R} \rightarrow \mathbb{R}$. We call 
\begin{equation}\label{eq:gAsConditionalExpectation}
\beta_g(x)=\frac{C(1-F(x_0))-\int_{x_0}^{x} g(u) f(u) du}{1-F(x)}\;.
\end{equation}

% \begin{equation}\label{eq:gAsConditionalExpectation}
% \beta_g(x)=\frac{\int_x^{\mathfrak{u}} g(t) f(t) dt}{1-F(x)}=\mathbb{E}(g(X)|X\geq x)
% \end{equation}
% is increasing and differentiable on the support of $X$.

Then, if $B=\beta_g(X)$, 
$$
h_{\beta_g}(x) = g(x) \text{ and } \beta_g(x_0)=C\;.
$$ 
If for some $t$, $x_0\leq t$ and $g$ is non-decreasing on $[x_0,t]$,  $\beta_g'(x)\geq (C-g(x))(1-F(x_0))f(x)/(1-F(x))$ for $x\in[x_0,t]$. Hence $\beta_g$ is increasing on $[x_0,t]$ if $g$ is non-decreasing and $g<C$.
%we have $h_{\beta_g}(x)= g(x)\;.$
%\begin{equation}\label{eq:SolnPhiBEqualsh}
% \psi_B(\beta_h(x_1))= h(x_1)\;.
% \end{equation}
\end{lemma}
% \begin{lemma}\label{lemma:keyODEs}
% Suppose $B=\genFunc(X_1)$, where $\genFunc$ is increasing and differentiable and $X_1$ is a random variable with cdf $F_1$ and pdf $f_1$. We denote by $\vValue_B$ the virtual value of the random variable $B$.  If $b=\genFunc(x_1)$, we have
% \begin{equation}\label{eq:ODEPhiG}
% \vValue_B(b)=\genFunc(x_1)-\genFunc'(x_1)\frac{1-F_1(x_1)}{f_1(x_1)}\;.
% \end{equation}
% Call for some $x_0$ and a function $h$
% $$
% \genFunc_h(x)=\frac{\genFunc_h(x_0)(1-F_1(x_0))-\int_{x_0}^{x} h(u) f_1(u) du}{1-F_1(x)}\;.
% $$
% If $B=\genFunc_h(X_1)$, we have $\vValue_B(\genFunc_h(x_1))= h(x_1)\;.$
% %\begin{equation}\label{eq:SolnPhiBEqualsh}
% % \vValue_B(\genFunc_h(x_1))= h(x_1)\;.
% % \end{equation}
% \end{lemma}
\begin{proof}
The result follows by simply differentiating the expression for $\beta_g$, and plugging-in the expression for $h_{\beta_g}$ obtained in Lemma \ref{definition_psi}. The result on the derivative is simple algebra.
\end{proof}
The two technical lemmas \ref{definition_psi} and \ref{lemma:keyODEs} show that for any non-decreasing function $g$, we can find a strategy $\beta_i$ such that the bid distribution induced by using $\beta_i$ on $F_{X_i}$ verifies $\psi_{B_i}(\beta_i(x)) = g(x)$ for all $x$ in the support of $F_{X_i}$, under Assumption \textbf{A2}.

%
% In Section \ref{subsec:thresholding_virtual_value}, we explained why sending a virtual value equal to zero when the initial one was negative increases the bidder's expected utility. To go from this virtual value to the corresponding bidding strategy $\beta$, we need to solve the simple ODE defined in Lemma \ref{definition_psi}. More formally, the following theorem shows how to improve any strategy assuming the bidder knows the current reserve price or value. This theorem works for non-regular value distributions and in the asymmetric case when the bidders have different value distributions.\nek{Conditions on $\psi$? This looks like an impossible result and will trigger referee backlash}
In Section \ref{subsec:thresholding_virtual_value}, we explained why sending to the seller a virtual value equal to zero when the initial one was negative increases the bidder's expected utility. To derive the corresponding bidding strategy $\beta$ from the virtual value, we need to solve the simple ODE defined in Lemma \ref{definition_psi}. More formally, Theorem \ref{thm:settingReserveValToZeroImprovesPerformance} shows how to improve any strategy assuming the bidder knows the current reserve price or value, which were computed to maximize bidders' payment using Myerson's Lemma above. This theorem still holds for non-regular value distributions and in the asymmetric case when the bidders have different value distributions.

\iffalse Intuitively, \redtext{for regular distributions}, there is no point for the bidder in having a virtual value that is negative somewhere since the seller sets the reserve price such that no bids corresponding to a negative virtual value will pass through the reserve price \nek{this reflects regular distribution thinking; needs a slight touch up I think}. 
\fi
\begin{figure}[t!]
\centering
\begin{tabular}{ccc}
\includegraphics[width=.40\linewidth]{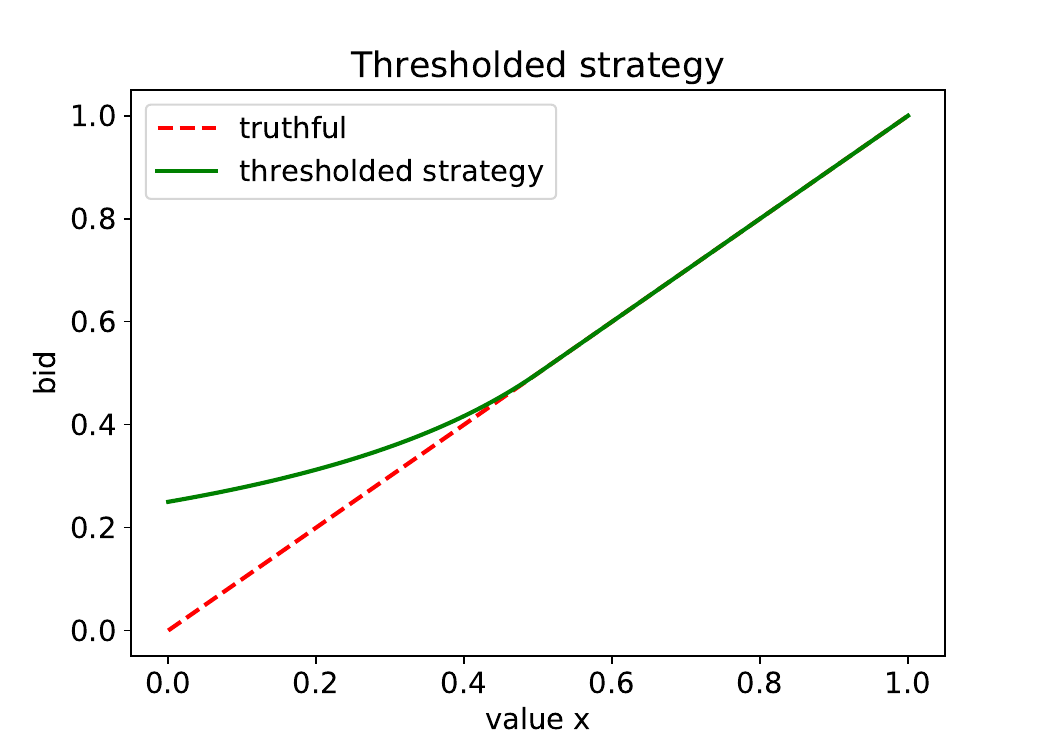} &
\includegraphics[width=.40\linewidth]{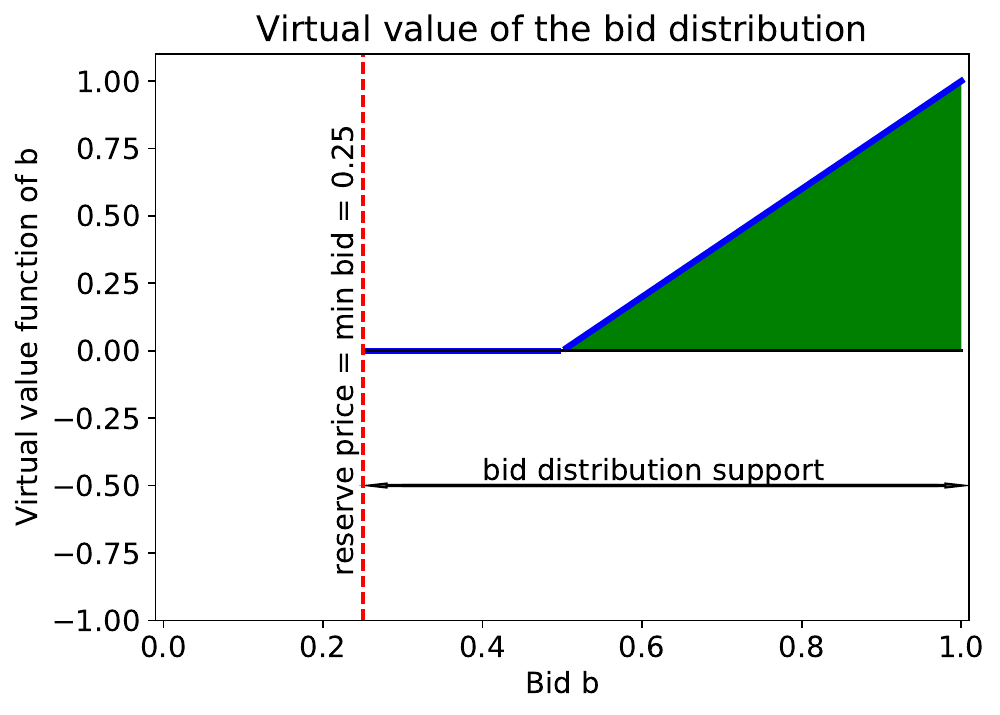}&

\end{tabular}	
\caption{\textbf{The value distribution is $\mathfrak{U}[0,1]$. Left: Thresholded strategy $\tilde{\beta}_{0.5}^{(0)}$ compared to the traditional truthful strategy. Right: virtual value of the bid distribution induced by the thresholded strategy.} The optimal reserve price of the thresholded strategy is equal to 0.25 (corresponding to a reserve value of 0) whereas the reserve price of the truthful strategy is equal to 0.5. (corresponding to a reserve value of 0.5). The green area represents the expected payment corresponding to the thresholded strategy (we assumed $G = 1$ for the sake of clarity).}
\label{fig:fig2} % I can do without the label too
\end{figure}
This improvement in bidder utility does not depend on the estimation of the competition and thus can easily be implemented in practice. We call this technique \emph{thresholding the virtual value at the monopoly price}.
We plot in Figure \ref{fig:fig2}, the bidding strategy $\tilde{\beta}_{0.5}^{(0)}$ corresponding to ``thresholding the virtual value at the monopoly price" - see formal definition in Equation \eqref{eq:DefNewThreshBidFunction} for $\newThreshNotation$ in general form -  when the strategic bidder's value distribution is $\mathfrak{U}[0,1]$ and the virtual value of the bid distribution induced by $\tilde{\beta}_{0.5}^{(0)}$ on $\mathfrak{U}[0,1]$. We recall that the monopoly price corresponding to $\mathfrak{U}[0,1]$ is equal to $0.5$. We remark that the strategy consists in overbidding below the monopoly price of the initial value distribution. The strategic bidder is ready to increase pointwise her payment when she wins auctions with low values in order to get a large decrease of the reserve price (going from $0.5$ to $0.25$). The minimum bid is equal to 0.25 and is equal to the reserve price. This explains why the expectation of the virtual value corresponding to the bid distribution is not equal to zero since the minimum bid is not zero.

Overall, the payment of the bidder remains unchanged compared to when the bidder was bidding truthfully with a reserve price equal to 0.5. Thresholding the virtual value at the monopoly price amounts to overbidding below the monopoly price, effectively providing over the course of the auctions an extra payment to the seller in exchange for lowering the reserve price/value faced by the strategic bidder. This strategy unlocks a very substantial utility gain for the bidder.

\subsection{Impact on bidder's utility}

Naturally, a key question is to understand the impact of this new strategy on the utility of the strategic bidder. We compare the situation with two bidders bidding truthfully against an optimal reserve price and the new situation with one bidder using the thresholded strategy and the second one bidding truthfully. We assume, as is standard in many textbooks and research papers numerical examples, that their value distribution is $\mathfrak{U}[0,1]$. 

Then, in this specific illustrative example, the strategic bidder utility increases from $1/12$ to $1/12 + (\log(2) - 1/2) / 4 \approx 0.132$ (\textbf{a 57\% increase}) and the welfare increases from 7/12 to $7/12 + (\log(2) - 1/2) / 4 \approx 0.632$. (\textbf{a 8\% increase}).

Using our technique, the strategic bidder improves their tradeoff between having a low reserve price and beating the competition compare to truthful bidding. The bidder could also underbid to decrease the reserve price. However, by underbidding, they would  decrease their probability to beat the competition in some situations. Using the same example, if the value distribution is $\mathfrak{U}[0,1]$, if the bidder divides their bid by two, they  also get a reserve price of 0.25. Yet, if there are two competitors with $\mathfrak{U}[0,1]$, this strategy incurs a decreases in utility from 0.057 (utility when truthful) to 0.041 (-28\%). Our strategy gets a utility 0.076 (+33\%) in the case of two competitors. Another fundamental difference is that our strategy, by overbidding, in the symmetric setting where all bidders have the same value distributions, increases the welfare of the system while underbidding strategies decreases it (with one competitor, if one bidder divides their bids by two, the welfare goes from 0.583 to 0.289 (-50\%). This is one reason explaining why the strategic bidder can get higher utility by overbidding~: they capture a large fraction of the resulting increase in welfare.

\iffalse These figures, which we computed analytically, can also be independently confirmed with a simple simulator playing auctions between a strategic bidder, a truthful bidder and a seller who is computing the reserve price of each bidder by maximizing $r_i(1-F_{B_i}(r_i))$. \nek{are we sounding too defensive with the simulator?}\fi 
In this example, as the initial reserve prices of the bidders are the same, we also remark that the utility of the truthful bidder and the global revenue of the seller remain unchanged.
\iffalse
\nek{I'm for cutting this paragraph}Furthermore, it costs nothing to the bidder employing the thresholded strategy to wait until the seller updates the current reserve price to the lowest of the monopoly prices of the thresholded strategies. Indeed, since the strategy only changes bids below the current reserve price, the strategic bidders pays nothing on average to try to convince the seller to decrease the reserve price. This strategy does not depend on any information about the competition (who is just assumed, as usual, to bid independently of the bidder we consider). 
\fi

To extend to other value distributions, with a log-normal distribution (which is widely used to model value distributions in online advertising) with parameters $\mu =  0.25$ and $\sigma = 1$, the utility of the strategic bidders goes from  0.791  to 1.025 (a  29.5\% increase). 

In this section, we introduced a simple strategy which guarantees a positive increase of utility for any possible distribution of the competition. In the next section, we show that by tuning a single parameter, i.e. the value where we threshold the virtual value, we can compute the optimal regularity-preserving strategy for a given distribution of the maximum bid of the competition.

\section{Optimal bidder's response given the distribution of the competition}
\label{sec:best_response_with_competition_knowledge}
%\section{Strategic improvements with increased side information}
The formulation introduced in the previous section offers a way to consider auction problems as functional-analytic problems and provides an alternative to game-theoretic methods used to derive optimal strategies and equilibria.  We now extend the result of the previous section to the case where the strategic bidder has access to the distribution of the highest bid of the competition. We show, for a specific given distribution of the competition, what is the optimal increasing and regularity-preserving (RP) strategy (Definition \ref{def:RP})). 

We impose the condition of being regularity-preserving since if the monopoly revenue curve function of the reserve price  has multiple local optima (in the case where the bid distribution is not regular), it requires in practice that the seller runs a global optimization procedure to compute all possible equivalent reserve prices. In practice, sellers are using various gradient descent algorithms \cite{paes2016field,shen2019learning}, forcing the bidder to use regularity-preserving strategies guaranteeing the seller converges to the right reserve price. We show in Section \ref{sec:numerical_approach} in some numerical experiments that when a bidding strategy implies two local optima of the payment curve, it does not make a difference for the seller but can change drastically bidder's utility.

We presented in the previous section a direct way to compute the expected utility of a bidding strategy $\beta_i$ when the seller is using a second price auction with personalized reserve price and the other bidders are bidding truthfully:
\begin{equation}\label{equ:utility_v2}
U(\beta_i) = \mathbb{E}_{X \sim F_{i}}\bigg((X-h_{\beta_i}(X))G_i(\beta(X))\textbf{1}(X \geq x_{\beta_i})\bigg)\;.
\end{equation}
with 
$h_{\beta_i}(x) = \beta_i(x) - \beta_i'(x)\frac{1-F_{X_i}(x)}{f_{X_i}(x)}\;$ and $x_{\beta_i} = h_{\beta_i}^{-1}(0)$.
In this section, we assume that the bidder has now access to the distribution of the highest bid of the competition that we denote by $G_i$ ($g_i$ being the associated pdf). 
% This form can be computed in many different settings and offers a way to compute optimal bidding strategies in some parametric class of functions by finding zeros of the directional derivatives.

In the following, unless otherwise stated, the expectation is taken according to the value distribution of the bidder. In order to be able to derive optimal strategies, we make use  of the previous expression and obtain the following lemma (we remove the subscript $i$ as it is now clear we consider bidder $i$).

We can optimize among the strategies with thresholded virtual values that we introduced in Section 2. 
We define more formally this class of bidding strategies. 

\begin{definition}[Thresholded bidding strategies]\label{definition_thresholded_strategies}
 A bidding strategy $\beta$ is called a thresholded bidding strategy if and only if there exists $r>0$ such that for all $x < r, h_{\beta}(x) = \psi_B(\beta(x)) = 0$. This family of functions can be parametrized with
\begin{equation*}
\beta_r^\gamma(x)=\frac{\gamma(r)(1-F(r))}{1-F(x)}\indicator{x< r}+\gamma(x)\indicator{x\geq r}\;,
\end{equation*}
with $r \in \mathbb{R}$ and $\gamma: \mathbb{R} \rightarrow \mathbb{R}$ continuous and increasing.
\end{definition}

This class of continuous bidding strategies has two degrees of freedom : the threshold $r$ such that for all $x < r, h_{\beta}(x) = 0$ and the strategy $\gamma$ used beyond the threshold. We do not restrict the functions  $\gamma$ that can be used beyond the threshold (beside being continuous and increasing). All the strategies defined in this class have the property that their reserve value is equal to zero, i.e. their reserve price is equal to their minimum bid, when the seller is welfare benevolent and the virtual value of $\gamma$ is positive beyond $r$. We first prove that the optimal regularity-preserving strategy belongs to the class of thresholded strategies. (We repeat that throughout the paper we assume that the strategic bidder draws values independently of the competition.)

\begin{lemma}\label{lemma:thresholdingIsABestResponse}
Let us consider $F_{i}$ regular (\textbf{A2} and \textbf{A3}) and denote by $x_0 = \inf(x : f(x) \neq 0)$ and  $x^* = \argmax(x(1-F(x)))$. Suppose further that (as implied by \textbf{A1} with $\eps>0$) $x(1-F(x))\tendsto 0$ as $x\tendsto c$ where $c$ is the right-end of the support of $F$ ($c$ could be infinite). If $F$ satisfies $x^* \neq x_0$, a best-response in the class of increasing regularity-preserving strategies belongs to the class of thresholded strategies. 
\label{opt_RP_lemma}
\end{lemma}
The formal proof is in Appendix. The proof is in several steps and exploits the property that $\psi_{B_i}$ has to be non-decreasing (since $\beta_i$ is regularity-preserving). We first show that, without loss of generality, the class of possible best-response strategies can be restricted to the class of strategies with non-negative $\psi_{B_i}$. Then, given $\beta$ a best-response in the class of of strategies satisfying the properties above, we show that there exists a thresholded strategy with at least the same utility. A similar lemma can be found in \cite{tang2016manipulate}. One main difference is that we place ourselves in the space of strategies instead of the space of bid distributions, which enables us to characterize precisely the required properties of the class of strategies considered and then, very importantly, to derive numerical methods for other types of auctions.   

We now show that there exists an optimal threshold $r$ for the strategic bidder that depends on the competition and that the optimal strategy to use for $x > r$ is to be truthful. We derive this result by computing the directional derivatives of the utility function defined in Equation \eqref{equ:utility_v2}.

\begin{theorem}\label{thm:quantificationOptimalThresholding}
Under the same assumptions as in Lemma \ref{lemma:thresholdingIsABestResponse}, a best response in the class of increasing and regularity-preserving strategies consists in thresholding at $r^*$ and bidding truthfully beyond $r^*$, where $r^*$ satisfies the equation, if $G$ is the distribution of the largest bid of the competition, 
\begin{equation*}\label{eq:keyEqThreshStratOneStrategic2}
G(r^*)=\mathbb{E}\bigg(\frac{X}{1-F(X)}g\left(\frac{r^*(1-F(r^*))}{1-F(X)}\right)\indicator{X\leq r^*}\bigg)\;.
\end{equation*}
\label{theorem2}
\end{theorem}
\begin{figure}[t]
\centering
\begin{tabular}{ccc}
\includegraphics[width=.33\linewidth]{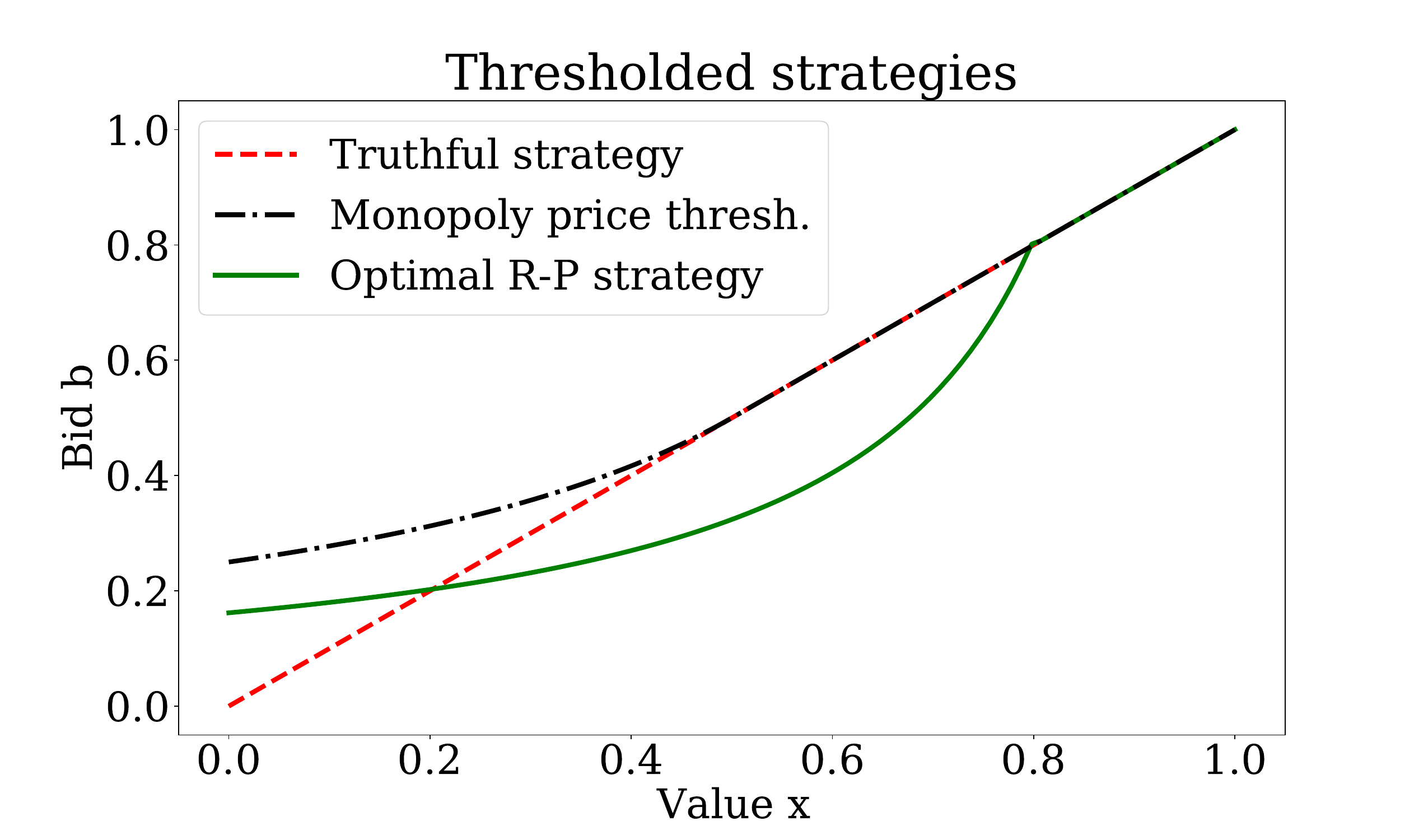} &
\includegraphics[width=.33\linewidth]{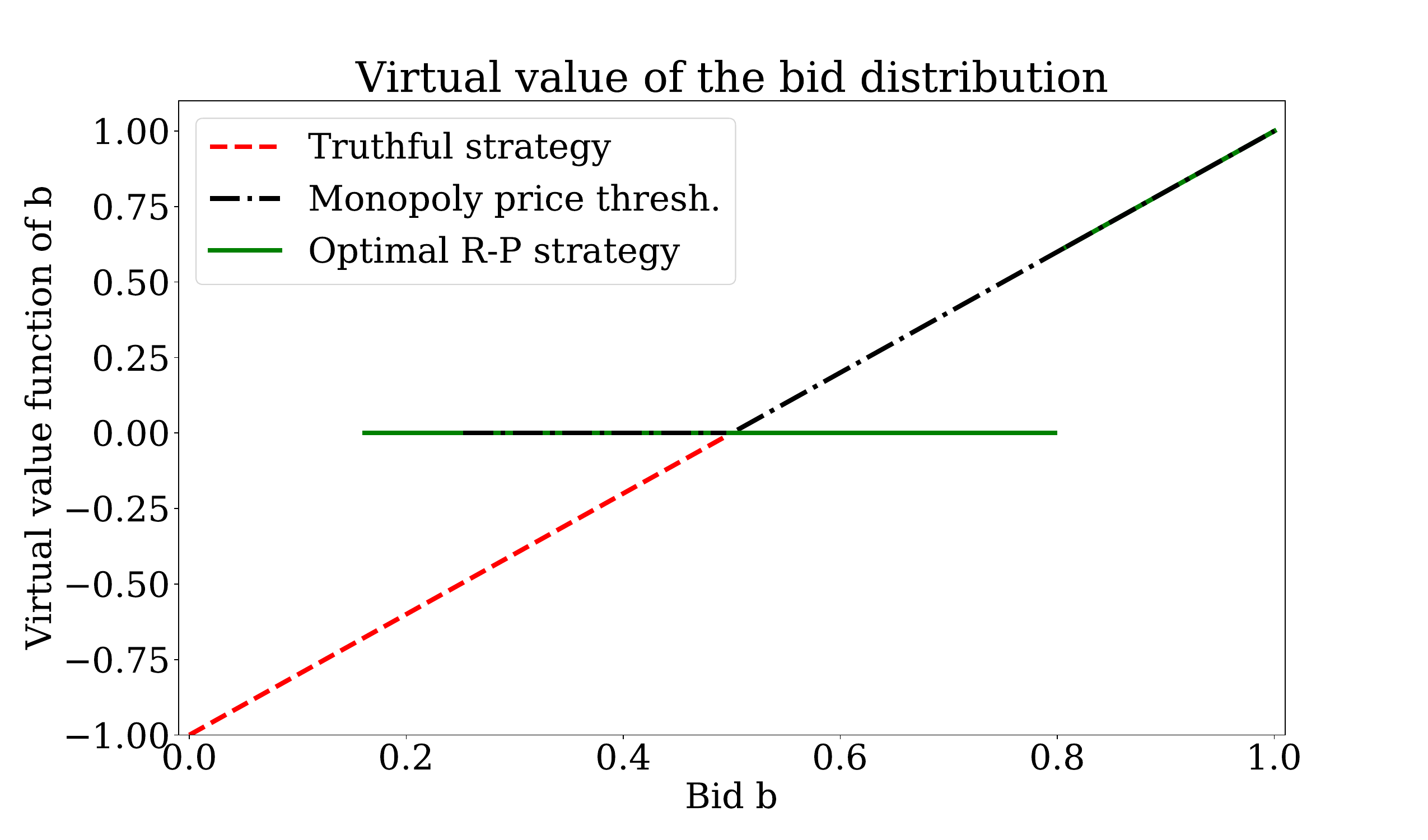}&
\includegraphics[width=.33\linewidth]{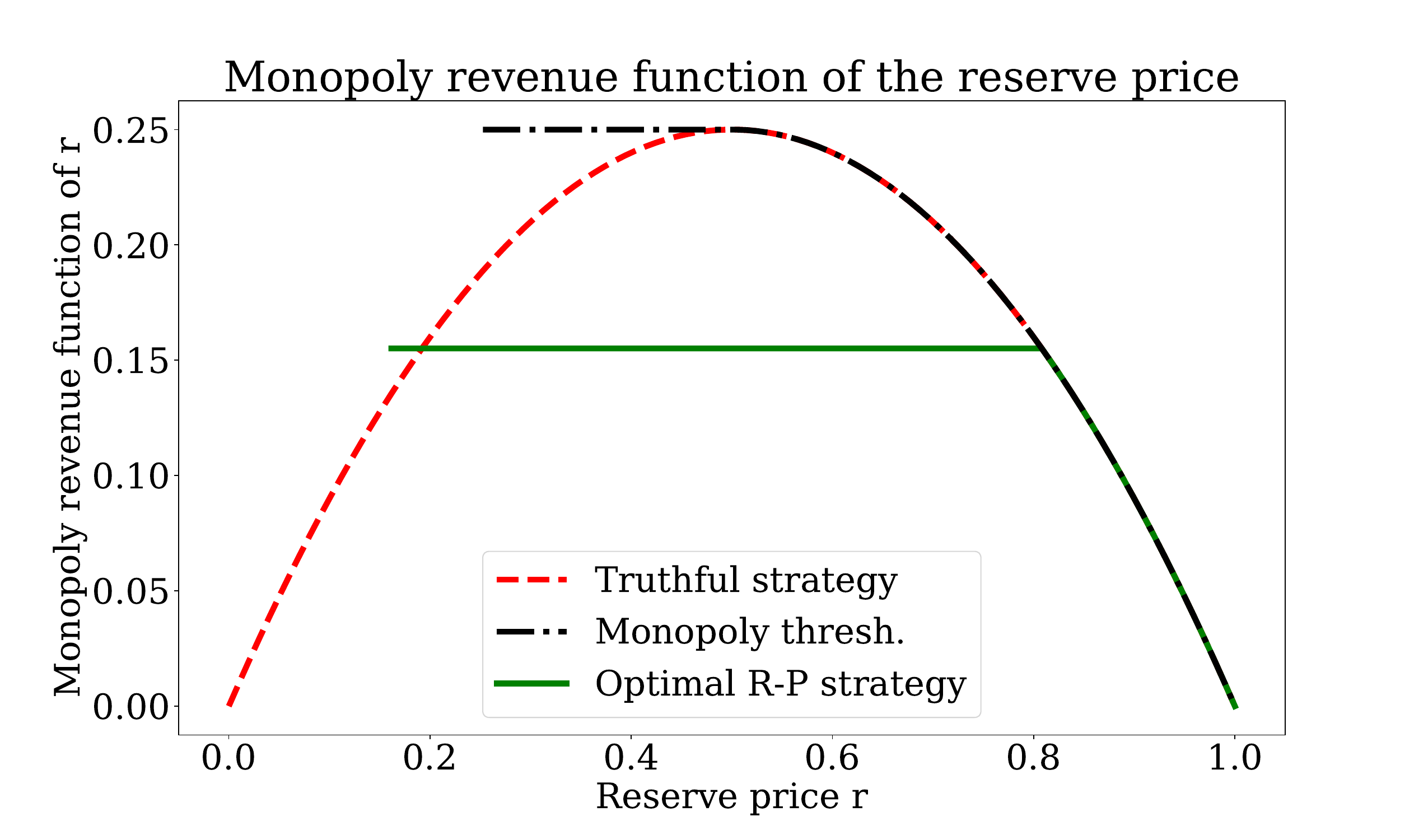} 

\end{tabular}	
\caption{\textbf{The value distribution is $\mathfrak{U}[0,1]$. Left: Thresholded strategy compared to the traditional truthful strategy. Middle: virtual value of the bid distribution induced by the thresholded strategy. Right: monopoly revenue of the induced bid distribution as function of the reserve price.}  }
\label{fig:fig3} % I can do without the label too
\end{figure}

In Theorem \ref{thm:settingReserveValToZeroImprovesPerformance}, we proved that when the strategic bidder does not know the distribution of the highest bid of the competition, he can use the thresholded strategy at his monopoly price and increases his utility compared to truthful bidding. Theorem \ref{thm:quantificationOptimalThresholding} gives the optimal threshold when the strategic bidder knows $G$. \iffalse These strategies are robust to local optimization of the seller since, if the seller only optimizes the reserve price locally, with the bidding strategies presented above, she has an incentive to decrease it at each step of the optimization. \nek{Unclear}\fi

\textbf{Some numerical results} We consider the situation where we have 1 strategic bidder, and 1 non-strategic one, both with $\mathfrak{U}[0,1]$ value distribution. We recall that the strategy introduced in Section \ref{sec:improvingAnyBiddingStrategy} was to bid truthfully beyond the monopoly price ($r=.5$ here) and using Theorem \ref{thm:settingReserveValToZeroImprovesPerformance} before.  This strategy yields a utility of $0.1316$, a $57\%$ increase over the standard truthful bidding revenue. The optimal strategy coming out of Theorem \ref{theorem2} consists in bidding truthfully beyond $r\simeq .8$ and using the thresholding completion before. The utility is then around 0.1468, a 76\% percent increase in bidder utility compared to bidding truthfully (truthful bidding yields a utility of $1/12\simeq .083$). This second strategy yields a higher utility for the strategic bidder but requires some knowledge of the competition. The optimal strategy in Theorem \ref{thm:quantificationOptimalThresholding} overbids on small values, underbids on intermediate values and is truthful on high values. We also recover that with no competition, the optimal strategy is to bid zero for any possible valuations. In Table \ref{table_intro}, we also notice that the difference in utility is decreasing with the number of players since, with increasing  competition, the strategic bidder can not lower his bid for values above his monopoly price. 
\section{Robustness to other strategic bidders}
\label{sec:bayes_nash}
%\subsection{Bayes-Nash (symmetric) equilibrium questions}
We have only discussed so far the situation where one bidder is strategic and the other bidders do not directly react to this strategy. It is a reasonable assumption in practice since the number of bidders able to implement sophisticated bidding strategies appears to be limited. We now investigate the case where all bidders are strategic. We assume that they all have the same value distribution $F_X$. Interestingly, we can analytically compute  what is the Nash equilibrium in the class of thresholded strategies introduced in Definition \ref{definition_thresholded_strategies}.

We can state a directional derivative result in this class of strategies by directionally-differentiating the expression of $U_i(\beta,...,\beta_t,...,\beta)$. As in Theorem \ref{thm:quantificationOptimalThresholding}, it implies the only strategy with 0 ``gradient" in this class is truthful beyond a value $r$ ($r$ is different from the one in Theorem \ref{thm:quantificationOptimalThresholding}), where $r$ can be determined through a non-linear equation. 
The proof is in Appendix. We can prove that there exists one unique symmetric Bayes-Nash equilibrium in the class of thresholded strategies which are robust to local optimization of the seller. In this equilibrium, bidders get the utility they receive in a second price auction without reserve price.

\begin{theorem}\label{thm:revBuyerInThreshStrategy}
We consider the symmetric setting where all the bidders have the same value distributions, satisfying Assumptions \textbf{A1} and \textbf{A2}. We furthermore assume for simplicity that the distribution is supported on [0,1]. Suppose this distribution has density that is continuous at 0 and 1 with $f(1)\neq 0$ - i.e. \textbf{A4} -  and $\psi_X$ crosses $0$ exactly once and is positive beyond that crossing point, i.e. \textbf{A3w}. There is a unique symmetric Nash equilibrium in the class of thresholded bidding strategies defined in Definition \ref{definition_thresholded_strategies}.
It is found by solving 
\begin{equation}\label{eq:keyEqThreshStratAllStrategic}
\frac{K-1}{r^*_{all}(1-F(r^*_{all}))}\mathbb{E}\bigg(XF^{K-2}(X)(1-F(X))\indicator{X\leq r^*_{all}}\bigg) = F^{K-1}(r^*_{all})\;.
\end{equation}
to determine $r^* (=r^*_{all})$ and bidding truthfully beyond $r^*$.

Moreover, if all the bidders are playing this strategy corresponding to the unique Nash equilibrium, the revenue of the seller is the same as in a second price auction with no reserve price. The same is true of the utility of the buyers. 	
	
\end{theorem}
The formal proof is in Appendix. 
%We recover some of the results of \cite{tang2016manipulate} but with a totally different approach. In their case, since they are using a non-constructive proof they are not able to exhibit the strategies corresponding to the Bayes-Nash equilibrium.

With appropriate shading functions, the bidders can recover the utility they would get when the seller was not optimizing her mechanism to maximize her revenue, a result independently found in \cite{tang2016manipulate} with a different proof.  
\section{Robustness to approximation error of the seller}
\label{sec:empirical_estimation}
In practice the seller needs to estimate the distribution of the buyer and hence does not have perfect knowledge of the bid distribution $F_{B_i}$. The buyer needs to find a robust shading method, making sure that the seller has an incentive to lower her reserve price, even if she misestimates the bid distribution. , If the seller is using a data-driven approach as in \cite{paes2016field} to set the reserve prices, we can still show that thresholding bidding strategies can improve the buyer’s utility.

We focus on the specific problem of empirical risk minimization (ERM)\citep{morgenstern2015pseudo,dhangwatnotai2015revenue,huang2018making}.

\begin{lemma}\label{lemma:impactOfMisestimationOnShading}
Suppose that the buyer uses a strategy $\beta$ under her value distribution $F$. Suppose the seller thinks that the value distribution of the buyer is $G$. 
Call $\lambda_F$ and $\lambda_G$ the hazard rate functions of the two distributions
Then the seller computes the virtual value function of the buyer under $G$, denoted $\psi_{B,G}$, as 
$$
\psi_{B,G}(\beta(x))=\psi_{B,F}(\beta(x))-\beta'(x)\left(\frac{1}{\lambda_G(x)}-\frac{1}{\lambda_F(x)}\right)\;.
$$	
\end{lemma}
As an aside, we note that by definition we have $\frac{1}{\lambda_G(x)}-\frac{1}{\lambda_F(x)}=\psi_F(x)-\psi_G(x)$. We have the following useful corollary pertaining to the thresholded strategies described in Section 3. 
\begin{corollary}\label{coro:impactOfMisestimationOnShadingThStrat}
If the buyer uses the strategy $\newThreshNotation(x)$ defined as
$$
\newThreshNotation(x)=\left(\frac{(r-\epsilon)(1-F(r))}{1-F(x)}+\epsilon\right)\indicator{x\leq r}+x\indicator{x>r}\;,
$$
we have, for $x\neq r$,
$\psi_{B,F}(\newThreshNotation(x))=\epsilon\indicator{x\leq r}+\psi_{F}(x)\indicator{x>r}\;.$
In particular, we have for $x\neq r$
$$
\left|\psi_{B,F}(\newThreshNotation(x))-\psi_{B,G}(\newThreshNotation(x)) \right|\leq \left|\psi_F(x)-\psi_G(x)\right|
 \left[(r-\epsilon)\indicator{x\leq r}+\indicator{x>r}\right]\;.
$$
If for all $x$, $\psi_{B,F}(\newThreshNotation)\geq \epsilon$ and $|\psi_F(x)-\psi_G(x)|\leq \delta$, we have 
$\psi_{B,G}(\newThreshNotation(x))\geq \epsilon -\delta\max((r-\epsilon),1)\;.
$
\end{corollary}
The previous results already give some results about the impact of empirically estimating the value distribution $F$ by the empirical cumulative distribution function $\hat{F}_n$ on setting the reserve price. However because our approximations are formulated in terms of hazard rate, applying those results would yield quite poor approximation results in the context of setting the monopoly price through ERM. This is due to the fact that estimating a density pointwise in supremum norm is a somewhat difficult problem in general, associated with poor rates of convergence. We refer the interested reader to \cite{TsybakovBook2009} for more details on this question. So we now focus on the specific problem of empirical minimization and take advantage of its characteristics to obtain better results than would have been possible by applying the results of the previous section. 
 
% Now more sketchy: the problem with above is that while estimating $F$ by $\hat{F}_n$ is usually accurate at speed $\sqrt{n}$ to estimate the inverse hazard rate we have to estimate the density in $\ell_\infty$ norm. My recollection from old results in a book by Tsybakov is that using kernel density estimators one can achieve that only at rate $\log(n)$ or poly$log(n)$. Of course the $\epsilon$ the buyer gives away is supposed to compensate the estimation error so if we have to go with estimation at speed $\log(n)$ it's really bad.

\begin{customthm}{3}\label{thm:epsAndMonopolyPriceByERM}
Suppose the buyer has a continuous and increasing value distribution $F$, supported on $[0,b]$, $b\leq \infty$, with the property that if $r\geq y\geq x$, $F(y)-F(x)\geq \gamma_F (y-x)$, where $\gamma_F>0$. Suppose that $\sup_{t\geq r}t(1-F(t))=r(1-F(r))$. Suppose the buyer uses the strategy $\newThreshNotation$ defined as
$$
\newThreshNotation(x)=\left(\frac{(r-\epsilon)(1-F(r))}{1-F(x)}+\epsilon\right)\indicator{x\leq r}+x\indicator{x>r}\;,
$$ Assume she samples $n$ values $\{x_i\}_{i=1}^n$ i.i.d according to the distribution $F$ and bids accordingly in second price auctions. Call $x_{(n)}=\max_{1\leq i \leq n}x_i$. In this case the (population) reserve value $x^*$ is equal to 0. Assume that the seller uses empirical risk minimization to determine the monopoly price in a (lazy) second price auction, using these $n$ samples. Call $\hat{x}_n^*$ the reserve value determined by the seller using ERM. We have, if $C_n(\delta)=n^{-1/2}\sqrt{\log(2/\delta)/2}$ and $\epsilon>x_{(n)} C_n(\delta)/F(r)$ with probability at least $1-\delta_1$, 
$$
\hat{x}_n^*<\frac{2rC_n(\delta)}{\epsilon\gamma_F} \text{ with probability at least } 1-(\delta+\delta_1) \;.
$$

In particular, if $\epsilon$ is replaced by a sequence $\epsilon_n$ such that $n^{1/2}\epsilon_n min(1,1/x_{(n)}) \rightarrow \infty$ in probability, 
$\hat{x}_n^*$ goes to 0 in probability like $n^{-1/2}\max(1,x_{(n)})/\epsilon_n$. 
\end{customthm}
Informally speaking, our theorem says that using the strategy $\newThreshNotationEpsn$  with $\epsilon_n$ slightly larger than $n^{-1/2}$ will yield a reserve value arbitrarily close to 0. Hence the population results we derived in earlier sections apply to the sample version of the problem. In future work, we plan to design games between bidders and seller where the seller can change the mechanism at any point and bidders can update their bidding strategy changing the bid distribution observed by the seller.  We also plan to consider the problem of the strategic reaction of the seller to the type of bidding strategies we have proposed in this paper. From an industrial standpoint, our work provides a new argument to come back to simple and more transparent auction mechanisms that are less subject to optimization on both the bidders' and the seller's sides. 

\section{Conclusion}
Reserve prices are learned in many practical situations using past bids. In this case, the celebrated second price auction is not anymore truthful. We propose an easy-to-implement strategy - dubbed \emph{``thresholding the virtual value at the monopoly price''} -  which keeps unchanged the expected payment of the strategic bidder, and increases very substantially her utility, even when the bidder has no information about the competition. This is possible as the strategic bidder overbids below the monopoly price, which can be interpreted as a form of payment to the seller in exchange for a lower reserve price. When all the bidders become strategic, we show that they can recover all the utility lost due to the introduction of reserve prices at a Nash equilibrium in the class of strategies we propose. We proved that our strategies are robust to various information structure of the game and to sample approximation errors.

In future work, we plan to design games between bidders and seller where the seller can change the mechanism at any point and bidders can update their bidding strategy changing the bid distribution observed by the seller.  We also plan to consider the problem of the strategic reaction of the seller to the type of bidding strategies we have proposed in this paper. From an industrial standpoint, our work provides a new argument to come back to simple and more transparent auction mechanisms that are less subject to optimization on both the bidders' and the seller's sides.

\bibliographystyle{ACM-Reference-Format}
\bibliography{bidderDependentAuctions}
\newpage

\section{Proof of results of Section 2}
\subsection{Proof of Lemma \ref{Myerson_lemma}}
\label{appendix_lemma1}
\begin{customlemma}{1}[Integrated version of the Myerson lemma]
Let bidder $i$ have value distribution $F_i$ and call $\beta_i$ her strategy, $F_{B_i}$ the induced distribution of bids and $\psi_{B_i}$  the corresponding virtual value function. Assume that  $F_{B_i}$ has a density and finite mean. Suppose that $i$'s bids are independent of the bids of the other bidders and denote by $G_i$ the cdf of the maximum of their bids. Suppose a lazy second price auction with reserve price denoted by $r$ is run. Then the payment $M_i$ of bidder $i$ to the seller can be expressed as 
\begin{equation*}
M(\beta_i) = \mathbb{E}_{B_i \sim F_{B_i}}\bigg(\psi_{B_i}(B_i)G_i(B_i)\textbf{1}(B_i \geq r)\bigg)\;.
\end{equation*}
When the other bidders are bidding truthfully, $G_i$ is the distribution of the maximum value of the other bidders.
\end{customlemma}
\begin{proof}
The proof is similar to the original one \cite{Myerson81} (see \cite{krishna2009auction} for more details).
It consists in using Fubini's theorem and integration by parts (this is why we need conditions on $F_{B_i}$)  to transform the standard form of the seller revenue, i.e.
$$
\mathbb{E}_{B_i\sim F_{B_i},B_j\sim F_{B_j}}\bigg(\max_{j\neq i}(B_j,r)\indicator{B_i\geq \max_{j\neq i}(B_j,r)}\bigg)
$$
into the above equation. 
We consider a lazy second price auction. Call $r$ the reserve price for bidder $i$. So the seller revenue from bidder $i$ is 
$$
\Exp{\max_{j\neq i}(B_j,r)\indicator{B_i\geq \max_{j\neq i}(B_j,r)}}
$$
In general, we could just call $Y_i=\max_{j\neq i}B_j$ and say that the revenue from $i$, or $i$'th expected payment is 
$$
\Exp{\max(Y_i,r)\indicator{B_i\geq \max(Y_i,r)}}\;.
$$
Call $G$ the cdf of $Y_i$ and $\tilde{G}$ the cdf of $\max(Y_i,r)$. Note that $\tilde{G}(t)=\indicator{t \geq r}G(t)$.

So if we note $B_i=t$, we have
$$
\Exp{\max(Y_i,r)\indicator{B_i\geq \max(Y_i,r)}|B_i=t}=\int_0^t u d\tilde{G}(u)
$$
Integrating by parts we get 
\begin{align*}
\int_0^t u d\tilde{G}(u)&=\left. u\tilde{G}(u)\right|_0^t-\int_0^t \tilde{G}(u)du\\
&=t\tilde{G}(t)-\int_0^t \tilde{G}(u)du=\indicator{t\geq r}\left[t G(t)-\int_{r}^t G(u)du\right]
\end{align*}
Hence, 
$$
\Exp{\max(Y_i,r)\indicator{b_i\geq \max(Y_i,r)}}=\int_0^\infty 
\left[\indicator{b_i\geq r}b_i G(b_i)-\int_{0}^{b_i}\indicator{u\geq r} G(u)du\right] f_{B_i}(b_i)db_i
$$
The first term of this integral is simply 
$$
\int_0^\infty \indicator{b_i\geq r}b_i G(b_i)f_{B_i}(b_i)db_i=
\Exp{B_iG(B_i)\indicator{B_i\geq r}}.
$$
Note that to split the two terms of the integral we need to assume that $\Exp{B_i}<\infty$, hence the first moment assumption on $F_i$.
The other part of the integral is 
\begin{align}
&\int_0^\infty \left(\int_{0}^{b_i}\indicator{u\geq r} G(u)du\right)f_{B_i}(b_i)db_i=\int_0^\infty \left(\int \indicator{b_i\geq u}\indicator{u\geq r} G(u)du\right)f_{B_i}(b_i)db_i\\
&=\int\int\indicator{u\geq r} G(u)  \indicator{b_i\geq u} f_{B_i}(b_i)  db_i du
=\int\indicator{u\geq r} G(u)  \left(\int\indicator{b_i\geq u} f(b_i)  db_i\right) du\\
&=\int\indicator{u\geq r} G(u)  P(B_i\geq u)du
= \int\indicator{u\geq r} G(u)  \frac{1-F_{B_i}(u)}{f_{B_i}(u)}f_{B_i}(u)du\\
&=\Exp{\indicator{B_i\geq r}G(B_i)\frac{1-F_{B_i}(B_i)}{f_{B_i}(B_i)}}
\end{align}
We used Fubini's theorem to change order of integrations, since all functions are non-negative. The result follows.

Of course, when $f_{B_i}(b_i)=0$ somewhere we understand $f_{B_i}(b_i)/f_{B_i}(b_i)=0/0$ as being equal to 1. To avoid this problem completely we can also simply write 
$$
M(\beta_i)=\int \left[b_if_{B_i}(b_i)-(1-F_{B_i}(b_i))\right]G(b_i)\indicator{b_i\geq r}db_i=\int \frac{\partial [b_i(F_{B_i}(b_i)-1)]}{\partial b_i}G(b_i) \indicator{b_i\geq x} db_i\;.
$$
If $F_{B_i}$ is not differentiable but absolutely continuous, its Radon-Nikodym derivative is used when interpreting the differentiation of $[b_i(F_{B_i}(b_i)-1)]$ with respect to $b_i$. 

\end{proof}
\subsection{Proof of Theorem 1}
This theorem works for non-regular value distributions and in the asymmetric case when the bidders have different value distributions.
\label{proof_theorem1}
\begin{customthm}{1}\label{thm:virtualvalue_onestrategic}
Consider the one-strategic setting in a lazy second price auction with $F_{X_i}$ the value distribution of the strategic bidder $i$ and a seller computing the reserve prices to maximize her revenue. No assumptions are needed on the distributions of other bidders. Suppose $\beta_r$ is a shading function with associated reserve value $r>0$. 
Then we can find $\tilde{\beta}_r$ such that: 
\textbf{1)} The reserve value associated with $\tilde{\beta}_r$ is 0.
\textbf{2)} $U(\tilde{\beta}_r)\geq U(\beta_r)$, $U$ being the utility of the buyer.
\textbf{3)} $R_i(\tilde{\beta}_r)\geq R_i(\beta_r)$, $R_i$ being the payment of bidder $i$ to the seller.
The following continuous functions fulfill these conditions for $\epsilon$ small enough:
$$
\tilde{\beta}^{(\epsilon)}_r(x)=\left(\frac{[\beta_r(r)-\epsilon](1-F_{X_i}(r))}{1-F_{X_i}(x)}+\epsilon\right)\textbf{1}_{x<r}+\beta_r(x)\textbf{1}_{x\geq r}
$$
\end{customthm}
\begin{proof}
The reserve value $r>0$ is given. 
Consider 
$$
\tilde{\beta}_r(x)=
\begin{cases}
t_r(x)& \text{ if } x\leq r\\
\beta_r(x) & 	\text{ if } x > r
\end{cases}
$$
To make things simple we require $t_r(r)=\beta_r(r^+)$, so we have continuity. 
Note that beyond $r$ the seller revenue is unaffected. 
If the seller sets the reserve value at $r_0$ the extra benefit compared to setting it at $r$ is 
$$
\Exp{\vValue_{t_r}(t_r(X)G(t_r(X))\indicator{r_0\leq X <r}}\;.
$$
Hence, as long as $\vValue_{t_r}(x)\geq 0$, the seller has an incentive to lower the reserve value. The extra gain to the buyer is 
$$
\Exp{(X-\vValue_{t_r}(t_r(X)))G(t_r(X))\indicator{r_0\leq X <r}}\;.
$$
Now, if we take 
$$
t_r(x)=\frac{t_r(0)}{1-F(x)}\;,
$$
it is easy to verify that 
$$
\vValue_{t_r}(t_r(x))=t_r(x)-t_r'(x) \frac{1-F(x)}{f(x)}=0\;.
$$
So in this limit case, there is no change in buyer's payment and when the reserve price is moved by the seller to any value on $[0,r]$. If we assume that the seller is welfare benevolent, she will set the reserve value to 0.  To have continuity of the bid function, we just require that 
$$
\frac{t_r(0)}{1-F(r)}=\beta_r(r^+)\;.
$$
Since there is no extra cost for the buyer, it is clear that his/her payoff is increased with this strategy. 
Taking $t_r^{(\eps)}$ such that 
$$
\vValue_{t_r^{(\eps)}}(t_r^{(\eps)}(x))=\eps\;,
$$
gives a strict incentive to the seller to move the reserve value to 0, (so the assumption that s/he is welfare benevolent is not required) even if it is slightly suboptimal for the buyer. 
Note that we explained in Lemma \ref{lemma:keyODEs} how to construct such a $\vValue$. In particular, 
$$
t_r^{\eps}=\frac{C_\eps}{1-F(x)}+\eps, \text{ with } \frac{C_\eps}{1-F(r)}+\eps=\beta_r(r^+)
$$
works.
Taking limits proves the result, i.e. for $\eps$ small enough the Lemma holds, since everything is continuous in $\eps$. 
\end{proof}
\textbf{Comment} We note that the flexibility afforded by $\eps$ is two-fold: when $\eps>0$, the extra seller revenue is a strictly decreasing function of the reserve price; hence even if for some reason reserve price movements are required to be small, the seller will have an incentive to make such move. The other reason is more related to estimation issues: if the reserve price is determined by empirical risk minimization, and hence affected by even small sampling noise, having $\eps$ big enough will guarantee that the mean extra gain of the seller will be above this sampling noise. Of course, the average cost for the bidder can be interpreted to just be $\eps$ at each value under the current reserve price and hence may not be a too hefty price to bear. 
\section{Proof of results of Section 3}
\subsection{Proof of Lemma 4}
\label{proof_lemma_4}
\begin{customlemma}{4}
Let us consider $F_{i}$ regular and denote by $x_0 = \inf(x : f(x) \neq 0)$ and  $x^* = \argmax(x(1-F(x)))$. Suppose further that $x(1-F(x)\tendsto 0$ as $x\tendsto \infty$. If $F$ verifies $x^* \neq x_0$, a best-response in the class of increasing regularity-preserving strategies belongs to the class of thresholded strategies. 
\end{customlemma}
\begin{proof}
The proof is in several steps and exploits the property that $\psi_{B_i}$ has to be non-decreasing (since $\beta_i$ is regularity-preserving). 
Let us prove that $x_\beta =\sup\{x : \psi_{B_i}(\beta(x)) < 0\} \leq x_0$. 

Assume $x_\beta >x_0$. Since $\psi_{B_i}$ is non-decreasing and $\beta$ increasing, $\forall x \in [x_0,x_\beta], \psi_{B_i}(\beta(x)) < 0$. Applying Theorem 1 on $[x_0,x_\beta]$, there exists $\hat{\beta}$ in the class of thresholded strategies such that $U(\hat{\beta}) \geq U(\beta)$. Hence, without loss of generality, the class of possible best-response strategies can be restricted to the class of strategies with non-negative $\psi_{B_i}$, whose corresponding reserve price is equal to their minimum bid.

We now focus on the class of strategies satisfying the properties above, a superset of the class of best response strategies. Since $\beta$ is regularity preserving, $x(1-F_{B_i}(x))$ is concave. Since $\psi_{B_i}$ is non-negative we also see that $x(1-F_{B_i}(x))$ is non-increasing. Therefore since $\beta$ is non-decreasing, we have that $\beta(x)(1-F_{B_i}(\beta(x)))$ is non-increasing by composition and therefore $\beta(x)(1-F_{i}(x))$ is non-increasing. 

Recall that $x_0 = \inf(x : f(x) \neq 0)$, the leftmost point in the support of $F$.  Let us consider $R = \beta(x_0)$. As the reserve price is the minimum bid, $R= \max_x(\beta(x)(1-F(x)))$. We call $R(x)=\beta(x)(1-F(x))$. As we explained above this function is non-increasing. 

We now show that there is a best-response in the class of thresholded strategy. We split the argument into two, depending on the value of $R$. 

$\bullet$ \textbf{Case $\mathbf{R > R_{truth}}$} Suppose  $R>R_{truth}$. We consider $ x_1= \inf\{x>a : R(x) \leq R_{truth}+(R - R_{truth})/2 = R'\}$. Note that $R'>R_{truth}=\max (x(1-F(x)))$, and therefore $\forall x, R'>x (1-F(x))$.

Note that if $x_1=x_0$, there is nothing to prove. Suppose $x_1>x_0$. Then by definition, $\beta(x)>R'/(1-F(x))$ on $[a,x_1]$. Let us replace on $[a,x_1]$, $\beta(x)$ by $\widetilde{\beta}(x) = R'/(1-F(x))$. $\widetilde{\beta}$ is still non-decreasing and since $R'>x(1-F(x))$, we also have $\beta(x)(1-F(x))>\widetilde{\beta}(x)(1-F(x))=R'>x(1-F(x))$. So on $[a,x_1]$, $\beta(x)>\widetilde{\beta}(x)>x$.  With this transformation, i.e. instead of using $\beta$, using the strategy $\widetilde{\beta}1_{x\leq x_1}+\beta \indicator{x>x_1}$ instead of $\beta$, we decrease the reserve price (going from R to R') and decrease the bid on $[a,x_1$], an interval on which the strategic bidder was bidding above his value pointwise. Hence, using standard arguments about truthful bidding being pointwise optimal from a utility maximization standpoint in second price auctions (with fixed reserves, \cite{krishna2009auction}, Proposition 2.1 p.13), we see that the utility of the strategic bidder has increased by moving from $\beta$ to $\widetilde{\beta}$, since the utility increased pointwise and the reserve has decreased. Hence, if $R>R_{truth}$, we have shown that we can improve our strategy by thresholding. Hence any strategy in this case can be improved by thresholding. We have also shown that no-best response strategy exists when $R>R_{truth}$.

%Hence we strictly increase bidder's utility and $R(a) \leq R_{truth}$.

$\bullet$ \textbf{Case $\mathbf{R \leq R_{truth}}$} We split the argument in two subcases. 

$\bullet$ \textbf{Case $\mathbf{x_0(1-F(x_0)) < R \leq R_{truth}}$} 
Let us define $x_1(R)$ by $x_1(R)=\inf \{x_1>x_0:x_1(1-F(x_1))=R\}$. Since $x(1-F(x))$ is concave and continuous $x_1(R)$ exists and $x_1(R)>x_0$. Assume $\beta$ is not thresholded on $[x_0,x_1]$. Let us now prove that $\beta(x)(1-F(x))$ and  $x(1-F(x))$ cross on $[x_0,x_1]$. We have $\beta(x_0)(1-F(x_0)) = R$, by definition of $R$. Using the fact that $\beta$ is not thresholded and $\beta(x)(1-F(x))$ is non-increasing, $\beta(x_1(R))(1-F(x_1(R))) < R=\beta(x_0)(1-F(x_0))$; also, by definition, $x_1(1-F(x_1)) = R$. 
As $\beta(x)(1-F(x))$ is non increasing on $[x_0,x_1(R)]$ and 
$x(1-F(x))$ is non decreasing on $[x_0,x_1(R)]$, if $R  \geq x_0$, there exists $x_1(R)\geq x_2>x_0$, s.t. 
$ \forall t \in [x_0,x_2],  \beta(t)(1-F(t)) \geq x_2(1-F(x_2)) \geq t(1-F(t))$. Let us pick one such $x_2$ and call $R'=x_2(1-F(x_2))$.
Consider the strategy 
$$
\widehat{\beta}(x)=\frac{R'}{1-F(x)}\indicator{x\leq x_2}+\beta(x)\indicator{x>x_2}\;.
$$
$\widehat{\beta}$ is increasing ($\beta(x_2)\geq R'$ by construction), has a lower reserve price ($R'$ instead of $R$) and decreases bids where $\beta$ was over-bidding compared to truthful bidding, i.e. on $[x_0,x_2]$. So it is at least as good as $\beta$ per our argument derived from \cite{krishna2009auction} above.

$\bullet$ \textbf{Case $\mathbf{R <  x_0(1-F(x_0))\leq R_{truth}}$} We consider two subcases, depending on whether $\beta(x)(1-F(x))$ and  $x(1-F(x))$ cross on the support of $F$ or not. Recall that $R=\beta(x_0)=\max_x \beta(x)(1-F(x))$, so in the case under consideration, $\beta(x_0)(1-F(x_0))<x_0(1-F(x_0))$ 

\textbf{a)} If $\beta(x)(1-F(x))$ and  $x(1-F(x))$ do not cross on the support of F, we have for all $x > x_0$, $\beta(x)(1-F(x)) \leq x(1-F(x))$.

Let us define $x_1(R)$ such that $x_1(R)=\inf \{x_1>x_0:x_1(1-F(x_1))=R\}$. By concavity and continuity of $x(1-F(x))$, and using the fact that $x(1-F(x))\tendsto 0$ at infinity, we have $x_1(R)$ exists and $x_1(R)>x_0$.

Furthermore, since $x(1-F(x))$ is concave, its super level-sets are convex so we have 
$\forall t \in [x_0,x_1(R)], t(1-F(t)) \geq R$. We also have $R=\sup_t \beta(t)(1-F(t))\geq \beta(t)(1-F(t))$ for all $t$. Hence, 
$
\forall t \in [x_0,x_1(R)],\; t(1-F(t)) \geq R \geq \beta(t)(1-F(t))\;.
$
Consider the strategy 
$$
\widehat{\beta}(x)=\frac{R}{1-F(x)}\indicator{x\leq x_1(R)}+x\indicator{x>x_1(R)}\;.
$$
We have $x \geq \hat{\beta}(x) \geq \beta(x)$.
So $\widehat{\beta}$ is increasing, has the same reserve price as $\beta$ and increases bids where $\beta$ was under-bidding compared to truthful bidding. So it is at least as good as $\beta$. 

\textbf{b)} If they cross, let $x_2=\inf\{x: \beta(x)\geq x \}$. Using the fact that $\beta(1-F)$ is non-increasing and $x(1-F)$ is concave, we have $ \forall t \in [x_0,x_2),  \beta(t)(1-F(t)) \leq x_2(1-F(x_2)) \leq  t(1-F(t))$.   By definition, we have $x_2\geq x_1(R)$, since $x_1(R)(1-F(x_1(R)))=R\geq \beta(t)(1-F(t))$ for all $t$ and $\beta(1-F)$ is non-increasing. 

Consider the strategy, if $x_2>x_1(R)$:
$$
\widehat{\beta}(x)=\frac{R}{1-F(x)}\indicator{x\leq x_1(R)} + x \indicator{x_2 \geq x > x_1(R)} + \beta(x)\indicator{x> x_2}\;.
$$
We have $x \geq \hat{\beta}(x) \geq \beta(x)$ on $[x_0,x_2]$.
So $\widehat{\beta}$ is non-decreasing (note that $\beta$ is non decreasing and $\widehat{\beta}(x_2)=x_2 \leq \beta(x)$ for $x>x_2$ by definition of $x_2$.) increasing, has the same reserve price  and increases bids where $\beta$ was under-bidding compared to truthful bidding. So it is at least as good as $\beta$ by the same arguments as before. 

If $x_2=x_1(R)$, and if $x_2=\min\{x: \beta(x)\geq x\}$, then $\beta(x_1(R))\geq x_1(R)$ and hence $\beta(x_1(R))(1-F(x_1(R))\geq x_1(R)(1-F(x_1(R))=R$. But $\beta(1-F)$ is non-increasing and $R=\sup_x \beta(x)(1-F(x))$. Hence we conclude that $\beta(x)(1-F(x))=R$ on $[x_0,x_1(R)]$ and hence $\beta$ is thresholded on $[x_0,x_1(R)]$ and there is nothing to show. 

Now suppose $x_2=x_1(R)$ is not attained; by definition of $x_2$,  we can find a decreasing sequence $y_n$, with $y_n\geq x_2=x_1(R)$ such that $\beta(y_n)\geq y_n$ and $y_n\tendsto x_1(R)=x_2$, otherwise $x_2$ would not be the infimum of the set we consider. Along this sequence, we have $\beta(y_n)(1-F(y_n))\geq y_n(1-F(y_n))$. We also have $R\geq \beta(y_n)(1-F(y_n))$, since $R\geq \beta(x)(1-F(x))$ for all $x$. Using the fact that $\beta(1-F)$ is non-increasing and $x(1-F(x))$ is continuous  to justify the existence of limits, we have 
$$
R\geq \lim_{n\tendsto \infty} \beta(y_n)(1-F(y_n))\geq \lim_{n\tendsto \infty} y_n(1-F(y_n))=x_1(R)(1-F(x_1(R)))=R\;.
$$
We conclude that $\lim_{n\tendsto \infty} \beta(y_n)(1-F(y_n))=R$. Using the fact $\beta(1-F)$ is non-increasing, we have $\beta(x_1(R))(1-F(x_1(R)))\geq \beta(y_n)(1-F(y_n)) \forall n$, since $x_1(R)\leq y_n$ and taking the limit (since $\beta(1-F)$ is non-increasing and $y_n$ is decreasing) we have 
$$
\beta(x_1(R))(1-F(x_1(R)))\geq \lim_{n\tendsto \infty}\beta(y_n)(1-F(y_n))=R\;.
$$
Since $R\geq \beta(x_1(R))(1-F(x_1(R)))$, we conclude that 
$$
R=\beta(x_1(R))(1-F(x_1(R)))\;, \text{ and by definition } R=x_1(R)(1-F(x_1(R)))\;.
$$
Hence $\beta(x_1(R))=x_1(R)$, $x_2$ is attained, $\beta$ is thresholded by our argument above and there is nothing to show.

\end{proof}

\subsection{Proof of Theorem \ref{theorem2}}
\label{appendix_section2}
\begin{customthm}{2}
Suppose that only one bidder is strategic, let G denote the CDF of the maximum value of the competition and g the corresponding pdf. Denote by $U(\beta_r^\gamma)$ the utility of the bidder using the strategy $\beta_r^\gamma$ according to the parametrization of Definition \ref{definition_thresholded_strategies}. Assume that the seller is welfare-benevolent. Then if $\beta_t=\beta_r^{\gamma+t\rho}$,
\begin{align*}
\left.\frac{\partial U(\beta_t)}{\partial t}\right|_{t=0}&=\mathbb{E}\bigg((X-\gamma(X))g(\gamma(X))\rho(X)\indicator{X\geq r}\bigg) 
\\&+\rho(r)(1-F(r))\bigg(\mathbb{E}\bigg(\frac{X}{1-F(X)}g\left(\frac{\gamma(r)(1-F(r))}{1-F(X)}\right)\indicator{X\leq r}\bigg)-G(\gamma(r))\bigg)
\end{align*}
\begin{align*}
\frac{\partial U(\beta_r^\gamma)}{\partial r}=-h_\beta(r)f(r)\bigg(\mathbb{E}\bigg(\frac{X}{1-F(X)}g\left(\frac{\gamma(r)(1-F(r))}{1-F(X)}\right)\indicator{X\leq r}\bigg)-G(\gamma(r))\bigg)
\end{align*}
The only strategy $\gamma$ and threshold $r$ for which we can cancel the derivatives in all directions $\rho$ consists in bidding truthfully beyond $r^*$, where $r^*$ satisfies the equation 
\begin{equation*}\label{eq:keyEqThreshStratOneStrategic}
G(r^*)=\mathbb{E}\bigg(\frac{X}{1-F(X)}g\left(\frac{r^*(1-F(r^*))}{1-F(X)}\right)\indicator{X\leq r^*}\bigg)\;.
\end{equation*}
\end{customthm}
\begin{proof}
Recall that our revenue in such a strategy (we take $\eps=0$) is just, when the seller is welfare-benevolent (and hence s/he will push the reserve value to 0 as long as $\vValue_B(\beta(x))\geq 0$ for all $x$)
$$
U(\beta_r^\gamma)=\Exp{(X-\vValue_B(\beta(X)))G(\beta(X))\indicator{X\geq r}}+\Exp{XG(\frac{\beta(r)(1-F(r))}{1-F(x)})\indicator{X\leq r}}\;.
$$
Now if $\beta_t=\beta+t \rho$, as usual, we have $\vValue_{B_t}(\beta_t)=h_\beta+t h_\rho$. Because we assumed that $\vValue_B(\beta(r))>0$, changing $\beta$ to $\beta_t$ won't drastically change that; in particular if $\vValue_{B_t}(\beta_t(r))>0$ the seller is still going to take all bids after $\beta_t(r)$.  In particular, we don't have to deal with the fact that the optimal reserve value for the seller may be completely different for $\beta$ and $\beta_t$. So the assumption $\vValue_B(\beta(r))>0$ is here for convenience and to avoid technical nuisances. In any case, we have 
\begin{align*}
\frac{\partial U(\beta_t)}{\partial t}&=\Exp{\left[-h_\rho(X)G(\beta(X))+(X-h_\beta(X))\rho(X)g(\gamma(X))\right]\indicator{X\geq r}}
\\&+\Exp{X\rho(r)\frac{1-F(r)}{1-F(X)}g(\frac{\gamma(r)(1-F(r))}{1-F(x)})\indicator{X\leq r}}\;.	
\end{align*}
Now using integration by parts on $\int_r^{\infty} \rho'(x)(1-F(x))G(\gamma(x))dx$, we have 
\begin{align*}
\Exp{-h_\rho(X)G(\beta(X))\indicator{X\geq r}}&=\left.\rho(x)(1-F(x))G(\gamma(x))\right|_{r}^\infty\\
&+\int_r^{\infty}[\rho(x)f(x)G(\gamma(x))-\rho(x)\beta'(x)g(\gamma(x))(1-F(x))]dx
\\&-\int_r^\infty \rho(x)f(x) G(\gamma(x))dx\\
&=-\rho(r)(1-F(r))G(\beta(r))-\Exp{\rho(X)\beta'(X)g(\gamma(X))\frac{1-F(X)}{f(X)}}\;.	
\end{align*}
Hence, 
\begin{gather*}
\frac{\partial U(\beta_t)}{\partial t}=\Exp{(X-\gamma(X))\rho(X)g(\gamma(X))\indicator{X\geq r}}-\rho(r)(1-F(r))G(\gamma(r))\\
+\rho(r)(1-F(r))\Exp{\left(\frac{X}{1-F(X)}\right)g(\frac{\gamma(r)(1-F(r))}{1-F(x)})\indicator{X\leq r}}\;.
\end{gather*}
This gives the first equation of Theorem \ref{theorem2}.

On the other hand, we have
\begin{gather*}
\frac{\partial U(\beta_r^\gamma)}{\partial r}=-(r-\vValue_B(\gamma(r)))G(\gamma(r))f(r)+rG(\gamma(r))f(r)\\
+\Exp{\frac{X}{1-F(X)}g(\frac{\gamma(r)(1-F(r))}{1-F(x)})\indicator{X\leq r}}\left[\gamma'(r)(1-F(r))-\beta(r)f(r)\right]\;.
\end{gather*}
Since 
$$
\left[\gamma'(r)(1-F(r))-\gamma(r)f(r)\right]=-f(r)\vValue_B(\beta(r))\;,
$$
we have established that 
\begin{gather*}
\frac{\partial U(\beta_r^\gamma)}{\partial r}=
\vValue_B(\gamma(r))f(r)\left[G(\gamma(r))-\Exp{\frac{X}{1-F(X)}g(\frac{\gamma(r)(1-F(r))}{1-F(x)})\indicator{X\leq r}}\right]\;.
\end{gather*}
We see that the only strategy $\beta$ and threshold $r$ for which we can cancel the derivatives in all directions $\rho$ consists in bidding truthfully beyond $r$, where $r$ satisfies the equation 
\begin{equation*}
G(r)=\Exp{\frac{X}{1-F(X)}g\left(\frac{r(1-F(r))}{1-F(X)}\right)\indicator{X\leq r}}\;.
\end{equation*}
Theorem \ref{theorem2} is shown.
\end{proof}
\section{Optimal non-robust best-response}
\label{sec:optimal_revenues}
\begin{figure}[H]
  \centering
  \includegraphics[width=0.30\textwidth]{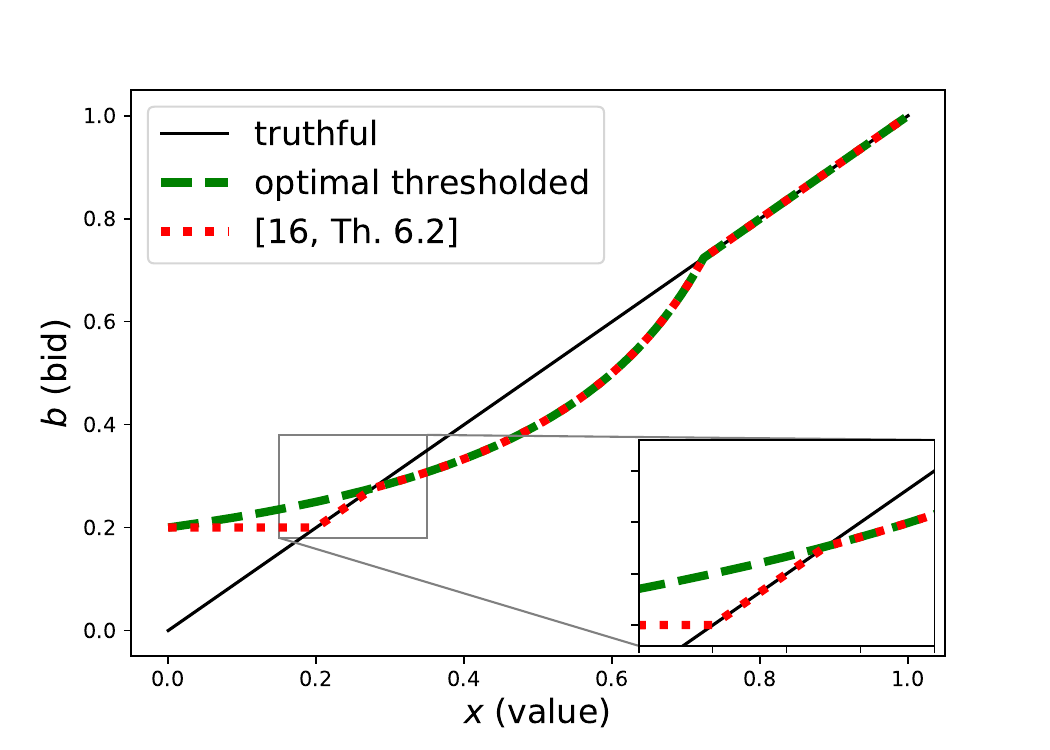}
  \includegraphics[width=0.30\textwidth]{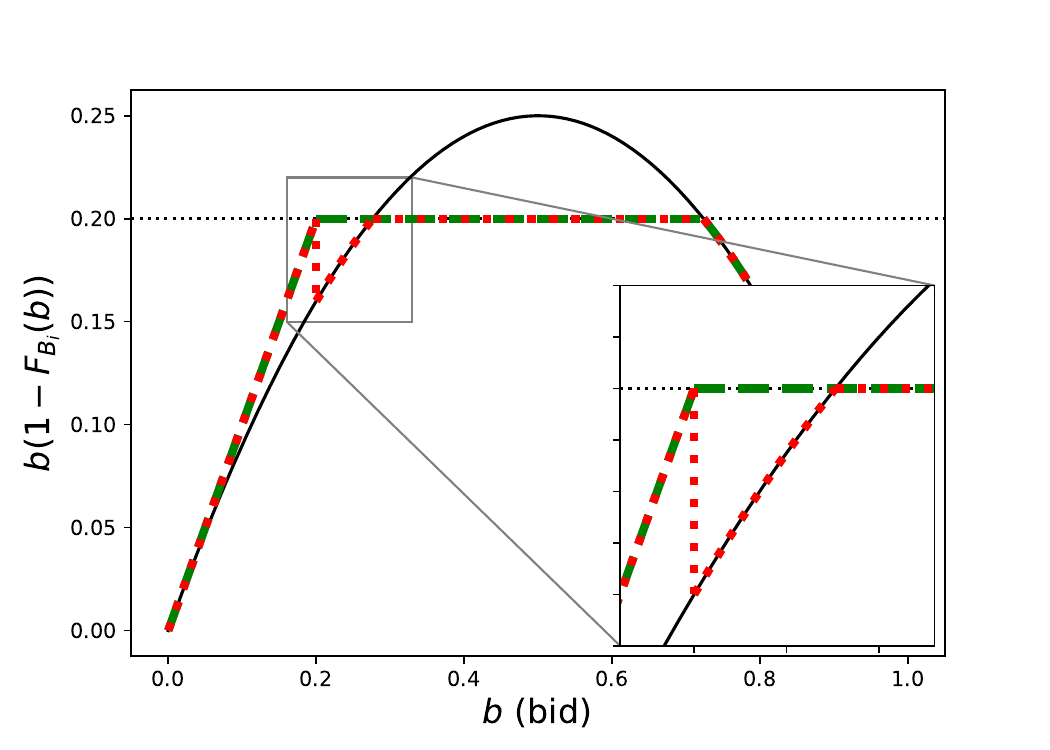}
  \caption{Left: illustration of strategies. Right: function optimized by the seller $b(1-F_{B}(b))$ depending on the strategy for values distribution $F$ being $\mathcal{U}(0,1)$.}
  \label{fig:optimal_revenues}
\end{figure}

 \cite[Th. 6.2]{tang2016manipulate} exhibits the best response for the one-strategic case. Unfortunately, this best response induces a complicated optimization problem for the seller: $R_{B}$ is non-convex with several local optima and the global optimum is reached at a discontinuity of $R_{B}$ (see Fig. \ref{fig:optimal_revenues}). This is especially problematic if sellers are known to optimize reserve prices conditionally on some available context using parametric models such as Deep Neural Networks \cite{dutting2017optimal} whose fit is optimized via first-order optimization and hence would regularly fail to find the global optimum. This is undesirable for the bidders: in any Stackelberg game, the leader's advantage comes from being able to predict the follower's strategy.

To address this issue, \cite{tang2016manipulate,nedelec2018thresholding} proposed a \emph{thresholding strategy} that is the best response in the restricted class of strategies that ensure $R_{B}$ to be concave as long as $R_{X}$ is so.

Figure \ref{fig:optimal_revenues} illustrates the different strategies as well as the corresponding optimization problems and the virtual values associated to the push-forward bid distributions. Understanding the strategic answer of the bidders helps avoiding worst-case scenario reasoning when studying the revenue of the sellers against strategic bidders. We now quantify precisely how the welfare is shared between strategic bidders and seller.
\newpage

\section{Proof of Section 4: deriving the Nash equilibrium}
\label{appendix_nash_equilibria}

\begin{lemma}
Consider the symmetric setting. Suppose that all the bidders are strategic and that they are all except one using the strategy $\beta$, $G_\beta$ denotes the CDF of the maximum bid of the competition when they use $\beta$ and $g_\beta$ the corresponding pdf, consider r and $\gamma$ such that $\beta = \beta_r^\gamma$. Assuming that the seller is welfare-benevolent and that $\beta_i^t=\beta_r^{\gamma+t\rho}$. Then  
\begin{align*}
&\left.\frac{\partial U(\beta,...,\beta_i^t,...,\beta)}{\partial t}\right|_{t=0}=\mathbb{E}\bigg((X-\gamma(X))g_{\beta}(\gamma(X))\rho(X)\indicator{X\geq r}\bigg) +\rho(r)\\
&(1-F(r))\bigg(\mathbb{E}\bigg(\frac{X}{1-F(X)}g_\beta\left(\frac{\gamma(r)(1-F(r))}{1-F(X)}\right)\indicator{X\leq r}\bigg)-G_\beta(\gamma(r))\bigg)
\end{align*}
Also, 
\begin{align*}
&\frac{\partial U(\beta,...,\beta_i,...,\beta)}{\partial r}=-h_\beta(r)f(r)\bigg(\mathbb{E}\bigg(\frac{X}{1-F(X)}g_\beta\left(\frac{\gamma(r)(1-F(r))}{1-F(X)}\right)\indicator{X\leq r}\bigg)-G_\beta(\gamma(r))\bigg)
\end{align*}
The only strategy $\gamma$ and threshold $r$ for which we can cancel/put to zero the derivatives in all directions $\rho$ consists in bidding truthfully beyond $r^*_{all}$, where $r^*_{all}$ satisfies the equation 
\begin{equation}\label{app:eq:keyEqThreshStratAllStrategic}
\frac{K-1}{r^*_{all}(1-F(r^*_{all}))}\mathbb{E}\bigg(XF^{K-2}(X)(1-F(X))\indicator{X\leq r^*_{all}}\bigg) = F^{K-1}(r^*_{all})\;.
\end{equation}
\end{lemma}
\begin{proof}
To derive the lemma we need to do the same calculations as in the previous case  but now the distribution of the highest bid of the competition $G_\beta$ depends on $\beta$. In the case of $K$ symmetric bidders with the same value distributions $F$, we have 
\begin{equation*}
G_\beta(x)= F^{K-1}(\beta^{-1}(x))\;\qquad\text{and}\qquad g_{\beta}(\beta(x))=\frac{(K-1)F^{K-2}(x)}{\beta'(x)}f(x)\;.
\end{equation*}
The last result follows by plugging-in these expressions in the corresponding equations.
\end{proof} 
\subsection{Proof of Theorem  \ref{thm:revBuyerInThreshStrategy}}
\label{appendix_theorem3}
\begin{customthm}{\ref{thm:revBuyerInThreshStrategy}}
We consider the symmetric setting where all the bidders have the same value distributions. We assume for simplicity that the distribution is supported on [0,1]. Suppose this distribution has density that is continuous at 0 and 1 with $f(1)\neq 0$ and X has a distribution for which $\psi_X$ crosses 0 exactly once and is positive beyond that crossing point. 

There is a unique symmetric Nash equilibrium in the class of thresholded bidding strategies defined in Definition \ref{definition_thresholded_strategies}.
It is found by solving Equation \eqref{app:eq:keyEqThreshStratAllStrategic} to determine $r^*$ and bidding truthfully beyond $r^*$

Moreover, if all the bidders are playing this strategy corresponding to the unique Nash equilibrium, the revenue of the seller is the same as in a second price auction with no reserve price. The same is true of the utility of the buyers. 	
	
\end{customthm}
Sketch of proof: 
\begin{itemize} 
\item State a directional derivative result in this class of strategies by directionally-differentiating the expression of $U_i(\beta,...,\beta_t,...,\beta)$.
\item Prove uniqueness of the Nash equilibrium.
\item Prove existence of the Nash equilibrium.
\item Show equivalence of revenue with a second price auction without reserve.
\end{itemize}

\subsection{A directional derivative result}
We can state a directional derivative result in this class of strategies by directionally-differentiating the expression of $U_i(\beta,...,\beta_t,...,\beta)$.It implies the only strategy with 0 ``gradient" in this class is truthful beyond a value $r$ ($r$ is different from the one in strategic case), where $r$ can be determined through a non-linear equation. 
\begin{lemma}\label{lemma:directionalDerivativesThresholdedStrategiesSymmetric}
Consider the symmetric setting. Suppose that all the bidders are strategic and that they are all except one using the strategy $\beta$, $G_\beta$ denotes the CDF of the maximum bid of the competition when they use $\beta$ and $g_\beta$ the corresponding pdf, consider r and $\gamma$ such that $\beta = \beta_r^\gamma$. Assuming that the seller is welfare-benevolent and that $\beta_i^t=\beta_r^{\gamma+t\rho}$. Then  
\begin{align*}
&\left.\frac{\partial U(\beta,...,\beta_i^t,...,\beta)}{\partial t}\right|_{t=0}=\mathbb{E}\bigg((X-\gamma(X))g_{\beta}(\gamma(X))\rho(X)\indicator{X\geq r}\bigg) +\rho(r)\\
&(1-F(r))\bigg(\mathbb{E}\bigg(\frac{X}{1-F(X)}g_\beta\left(\frac{\gamma(r)(1-F(r))}{1-F(X)}\right)\indicator{X\leq r}\bigg)-G_\beta(\gamma(r))\bigg)
\end{align*}
Also, 
\begin{align*}
&\frac{\partial U(\beta,...,\beta_i,...,\beta)}{\partial r}=-h_\beta(r)f(r)\bigg(\mathbb{E}\bigg(\frac{X}{1-F(X)}g_\beta\left(\frac{\gamma(r)(1-F(r))}{1-F(X)}\right)\indicator{X\leq r}\bigg)-G_\beta(\gamma(r))\bigg)
\end{align*}
The only strategy $\gamma$ and threshold $r$ for which we can cancel/put to zero the derivatives in all directions $\rho$ consists in bidding truthfully beyond $r^*_{all}$, where $r^*_{all}$ satisfies the equation 
\begin{equation}\label{app:eq:keyEqThreshStratAllStrategicReStatement}
\frac{K-1}{r^*_{all}(1-F(r^*_{all}))}\mathbb{E}\bigg(XF^{K-2}(X)(1-F(X))\indicator{X\leq r^*_{all}}\bigg) = F^{K-1}(r^*_{all})\;.
\end{equation}
\end{lemma}
\begin{proof}
In the case of $K$ symmetric bidders with the same value distributions $F$, we have 
\begin{equation*}
G_\beta(x)= F^{K-1}(\beta^{-1}(x))\;\qquad\text{and}\qquad g_{\beta}(\beta(x))=\frac{(K-1)F^{K-2}(x)}{\beta'(x)}f(x)\;.
\end{equation*}
The last result follows by plugging-in these expressions in the corresponding equations.
\end{proof} 

\subsection{Uniqueness of the Nash equilibrium}
We show that this strategy represents the unique symmetric Bayes-Nash equilibrium in the class of shading functions defined in Definition \ref{definition_thresholded_strategies}. At this equilibrium, the bidders recover the utility they would get in a second price auction without reserve price.

\begin{lemma}\label{lemma:keyEqThreshAllStratHasUniqueSolution}
Suppose $X$ has a distribution for which $\psi_X$ crosses $0$ exactly once and is positive beyond that crossing point. Then Equation \eqref{app:eq:keyEqThreshStratAllStrategicReStatement} has a unique non-zero solution. 
\end{lemma}
\begin{proof}
We have by integration by parts  
\begin{align*}
&(K-1)\Exp{\bigg(XF^{K-2}(X)(1-F(X))\indicator{X\leq r}\bigg)}
\\&=r(1-F(r))F^{K-1}(r)+\Exp{\bigg(\vValue_X(X) F^{K-1}(X)\indicator{X\leq r}\bigg)}\;.
\end{align*}
Hence finding the root of 
$$
\frac{K-1}{r(1-F(r))}\Exp{\bigg(XF^{K-2}(X)(1-F(X))\indicator{X\leq r}\bigg)} = F^{K-1}(r)
$$
amounts to finding the root(s) of 
$$
\mathfrak{R}(r)\triangleq\Exp{\bigg(\vValue_X(X) F^{K-1}(X)\indicator{X\leq r}\bigg)}=0\;.
$$
0 is an obvious root of the above equation but does not work for the penultimate one.
Note that for the class of distributions we consider (which is much larger than regular distributions but contains it), this function $\mathfrak{R}$ is decreasing and then increasing after $\vValue_X^{-1}(0)$, since the virtual value is negative and then positive. Since $\mathfrak{R}(0)=0$, it will have at most one non-zero root for the distributions we consider. The fact that this function is positive at infinity (or at the end of the support of $X$) comes from the fact that its value then is the revenue-per-buyer of a seller performing a second price auction with $K$ symmetric buyers bidding truthfully with a reserve price of 0. And this is by definition positive. 
So we have shown that for regular distributions and the much broader class of distributions we consider the function $\mathfrak{R}$ has exactly one non-zero root. 
\end{proof} 
\subsection{Existence of the Nash equilibrium}
\subsubsection{Best response strategy: one strategic case} 
\begin{lemma}
Suppose that only one bidder is strategic, let G denote the CDF of the maximum value of the competition and g the corresponding pdf. Denote by $U(\beta_r^\gamma)$ the utility of the bidder using the strategy $\beta_r^\gamma$ according to the parametrization of Definition \ref{definition_thresholded_strategies}. Assume that the seller is welfare-benevolent. Then if $\beta_t=\beta_r^{\gamma+t\rho}$,
\begin{align*}
&\left.\frac{\partial U(\beta_t)}{\partial t}\right|_{t=0}=\mathbb{E}\bigg((X-\gamma(X))g(\gamma(X))\rho(X)\indicator{X\geq r}\bigg) +\rho(r)\\
&(1-F(r))\bigg(\mathbb{E}\bigg(\frac{X}{1-F(X)}g\left(\frac{\gamma(r)(1-F(r))}{1-F(X)}\right)\indicator{X\leq r}\bigg)-G(\gamma(r))\bigg)
\end{align*}
\begin{align*}
\frac{\partial U(\beta_r^\gamma)}{\partial r}=&-h_\beta(r)f(r)
\\&\bigg(\mathbb{E}\bigg(\frac{X}{1-F(X)}g\left(\frac{\gamma(r)(1-F(r))}{1-F(X)}\right)\indicator{X\leq r}\bigg)-G(\gamma(r))\bigg)
\end{align*}
The only strategy $\gamma$ and threshold $r$ for which we can cancel/put to zero the derivatives in all directions $\rho$ consists in bidding truthfully beyond $r^*$, where $r^*$ satisfies the equation 
\begin{equation}\label{eq:keyEqThreshStratOneStrategic}
G(r^*)=\mathbb{E}\bigg(\frac{X}{1-F(X)}g\left(\frac{r^*(1-F(r^*))}{1-F(X)}\right)\indicator{X\leq r^*}\bigg)\;.
\end{equation}
\label{lemma:directionalDerivativesThresholdedStrategiesRight}
\end{lemma}
\begin{proof} 
Recall that our revenue in such a strategy (we take $\eps=0$) is just, when the seller is welfare-benevolent (and hence s/he will push the reserve value to 0 as long as $\vValue_B(\beta(x))\geq 0$ for all $x$)
$$
U(\beta_r^\gamma)=\Exp{(X-\vValue_B(\beta(X)))G(\beta(X))\indicator{X\geq r}}+\Exp{XG(\frac{\beta(r)(1-F(r))}{1-F(x)})\indicator{X\leq r}}\;.
$$
Now if $\beta_t=\beta+t \rho$, as usual, we have $\vValue_{B_t}(\beta_t)=h_\beta+t h_\rho$. Because we assumed that $\vValue_B(\beta(r))>0$, changing $\beta$ to $\beta_t$ won't drastically change that; in particular if $\vValue_{B_t}(\beta_t(r))>0$ the seller is still going to take all bids after $\beta_t(r)$.  In particular, we don't have to deal with the fact that the optimal reserve value for the seller may be completely different for $\beta$ and $\beta_t$. So the assumption $\vValue_B(\beta(r))>0$ is here for convenience and to avoid technical nuisances. In any case, we have 
\begin{gather*}
\frac{\partial U(\beta_t)}{\partial t}=\Exp{\left[-h_\rho(X)G(\beta(X))+(X-h_\beta(X))\rho(X)g(\gamma(X))\right]\indicator{X\geq r}}\\
+\Exp{X\rho(r)\frac{1-F(r)}{1-F(X)}g(\frac{\gamma(r)(1-F(r))}{1-F(x)})\indicator{X\leq r}}\;.	
\end{gather*}
Now using integration by parts on $\int_r^{\infty} \rho'(x)(1-F(x))G(\gamma(x))dx$, we have 
\begin{align*}
\Exp{-h_\rho(X)G(\beta(X))\indicator{X\geq r}}&=\left.\rho(x)(1-F(x))G(\gamma(x))\right|_{r}^\infty\\
&+\int_r^{\infty}[\rho(x)f(x)G(\gamma(x))-\rho(x)\beta'(x)g(\gamma(x))(1-F(x))]dx\\
&-\int_r^\infty \rho(x)f(x) G(\gamma(x))dx\\
&=-\rho(r)(1-F(r))G(\beta(r))-\Exp{\rho(X)\beta'(X)g(\gamma(X))\frac{1-F(X)}{f(X)}}\;.	
\end{align*}
Hence, 
\begin{gather*}
\frac{\partial U(\beta_t)}{\partial t}=\Exp{(X-\gamma(X))\rho(X)g(\gamma(X))\indicator{X\geq r}}-\rho(r)(1-F(r))G(\gamma(r))\\
+\rho(r)(1-F(r))\Exp{\left(\frac{X}{1-F(X)}\right)g(\frac{\gamma(r)(1-F(r))}{1-F(x)})\indicator{X\leq r}}\;.
\end{gather*}
This gives the first equation of Lemma \ref{lemma:directionalDerivativesThresholdedStrategiesRight}.

On the other hand, we have
\begin{gather*}
\frac{\partial U(\beta_r^\gamma)}{\partial r}=-(r-\vValue_B(\gamma(r)))G(\gamma(r))f(r)+rG(\gamma(r))f(r)\\
+\Exp{\frac{X}{1-F(X)}g(\frac{\gamma(r)(1-F(r))}{1-F(x)})\indicator{X\leq r}}\left[\gamma'(r)(1-F(r))-\beta(r)f(r)\right]\;.
\end{gather*}
Since 
$$
\left[\gamma'(r)(1-F(r))-\gamma(r)f(r)\right]=-f(r)\vValue_B(\beta(r))\;,
$$
we have established that 
\begin{gather*}
\frac{\partial U(\beta_r^\gamma)}{\partial r}=
\vValue_B(\gamma(r))f(r)\left[G(\gamma(r))-\Exp{\frac{X}{1-F(X)}g(\frac{\gamma(r)(1-F(r))}{1-F(x)})\indicator{X\leq r}}\right]\;.
\end{gather*}
We see that the only strategy $\beta$ and threshold $r$ for which we can cancel/put to zero the derivatives in all directions $\rho$ consists in bidding truthfully beyond $r$, where $r$ satisfies the equation 
\begin{equation*}
G(r)=\Exp{\frac{X}{1-F(X)}g\left(\frac{r(1-F(r))}{1-F(X)}\right)\indicator{X\leq r}}\;.
\end{equation*}
Lemma \ref{lemma:directionalDerivativesThresholdedStrategiesRight} is shown.
\end{proof} 
\subsubsection{Proof of existence of the Nash equilibrium}
$\bullet$ \textbf{On Equation \eqref{eq:keyEqThreshStratOneStrategic} and consequences}

Recall the statement of Equation \eqref{eq:keyEqThreshStratOneStrategic}. 
\begin{equation}\label{eq:keyeqOneStratExistence}
\Exp{\frac{X}{1-F(X)}g\left(\frac{r(1-F(r))}{1-F(X)}\right)\indicator{X\leq r}}=G(r)\;.
\tag{\ref{eq:keyEqThreshStratOneStrategic}}
\end{equation}
\begin{lemma}\label{lemma:atMostOneNonTrivialSoln}
Suppose $X$ has a regular distribution or a distribution whose virtual value crosses 0 exactly once and is negative to the left of this crossing point and positive to the right of that point. Suppose further that this distribution is compactly supported (on [0,1] for convenience). 	
Equation  \eqref{eq:keyeqOneStratExistence} can be re-written as 
\begin{equation}\label{eq:keyeqOneStratExistenceAltForm}
\Exp{\frac{X}{1-F(X)}g\left(\frac{r(1-F(r))}{1-F(X)}\right)\indicator{X\leq r}}=G(r)+
\frac{1}{r(1-F(r))}\Exp{\vValue_X(X)G\left(\frac{r(1-F(r))}{1-F(X)}\right)\indicator{X\leq r}}\;.
\end{equation}
Equation \eqref{eq:keyeqOneStratExistence}	has at most one solution on (0,1). This possible root is greater than $\vValue^{-1}(0)$. 0 is also a (trivial) solution of  Equation \eqref{eq:keyeqOneStratExistence}.
\end{lemma}
The assumption that $X$ is supported on [0,1] can easily be replaced by the assumption that it is compactly supported, but it made notations more convenient. 
\begin{proof}[Proof of Lemma \ref{lemma:atMostOneNonTrivialSoln}]
We use again integration by parts~:
\begin{gather*}
\Exp{\frac{X}{1-F(X)}g\left(\frac{r(1-F(r))}{1-F(X)}\right)\indicator{X\leq r}}=
\int_0^r\frac{x(1-F(x))}{(1-F(x))^2}g\left(\frac{r(1-F(r))}{1-F(x)}\right)f(x) dx\\
=\left. x(1-F(x))\frac{G\left(\frac{r(1-F(r))}{1-F(x)}\right)}{r(1-F(r))}\right|_0^r
-\frac{1}{r(1-F(r))}\int_0^r (x(1-F(x)))' G\left(\frac{r(1-F(r))}{1-F(x)}\right) dx\\
=G(r)+\frac{1}{r(1-F(r))}\Exp{\vValue_X(X)G\left(\frac{r(1-F(r))}{1-F(X)}\right)\indicator{X\leq r}}\;.
\end{gather*}
So we are really looking at the properties of the solution of 
$$
\frac{1}{r(1-F(r))}\Exp{\vValue_X(X)G\left(\frac{r(1-F(r))}{1-F(X)}\right)\indicator{X\leq r}}=0\;.
$$
We call 
$$
I(r)=\Exp{\vValue_X(X)G\left(\frac{r(1-F(r))}{1-F(X)}\right)\indicator{X\leq r}}\;.
$$
For regular distributions, or distribution for which $\vValue_X$ crosses 0 exactly once and is negative to the left of this crossing point and positive to the right of it, it is clear that if $r=\vValue^{-1}(0)$, $I(r)<0$.  

Now we note that 
$$
\frac{\partial I(r)}{\partial r}=\vValue_X(r)G\left(\frac{r(1-F(r))}{1-F(r)}\right)-\vValue_X(r)\Exp{\frac{\vValue_X(X)}{1-F(X)}g\left(\frac{r(1-F(r))}{1-F(X)}\right)\indicator{X\leq r}}\;.
$$
Using $\vValue_X(x)=x-(1-F(x))/f(x)< x$, we see that 
$$
\Exp{\frac{\vValue_X(X)}{1-F(X)}g\left(\frac{r(1-F(r))}{1-F(X)}\right)\indicator{X\leq r}}< \Exp{\frac{X}{1-F(X)}g\left(\frac{r(1-F(r))}{1-F(X)}\right)\indicator{X\leq r}}
$$
So if $r$ is a solution of 
$$
\Exp{\frac{X}{1-F(X)}g\left(\frac{r(1-F(r))}{1-F(X)}\right)\indicator{X\leq r}}=G(r)
$$
we have, if $r>\vValue_X^{-1}(0)$ and $r\neq 1$, 
$$
\frac{\partial I(r)}{\partial r}>0\;.
$$
So $r$ needs to be a solution of $I(r)=0$ (which is equivalent to the initial equation for non-trivial solutions) and must have $\frac{\partial I(r)}{\partial r}>0$.

So we have shown that $I$ is a function such that its (non-trivial) zeros are such that $I$ is strictly increasing at those roots. Because $I$ is differentiable and hence continuous, this implies that $I$ can have at most one non-trivial root. (0 is a trivial root of $I(r)=0$, though it is not an acceptable solution of our initial problem.)

We note that the end point of the support of $X$ is also a trivial solution of $I(r)=0$, by the dominated convergence theorem, though not an acceptable solution of our initial problem, as shown by a simple inspection.

\end{proof} 

\begin{lemma}\label{lemma:uniqueSolnInOneStrategicCase}
Suppose that $G(0)=0$, $G$ is continuous and either $G(x)=g^{k}(0) x^k+o(x^k)$ near 0 for some $k$ or $G$ is constant near 0.  Assume $f$ has a continuous density near 1 with $f(1)\neq 0$ and the Assumptions of Lemma \ref{lemma:atMostOneNonTrivialSoln} are satisfied. Then Equation 
$$
\mathfrak{G}(r)=\frac{1}{r(1-F(r))}\Exp{\vValue_X(X)G\left(\frac{r(1-F(r))}{1-F(x)}\right)\indicator{X\leq r}}=0
$$
has a unique root in $(0,1)$. 

In particular in this situation there exists an optimal strategy in the class of shading functions defined in Definition \ref{definition_thresholded_strategies} and it is unique. It is defined by being truthful beyond $r^*: \mathfrak{G}(r^*)=0$ and shading in such a way that our virtual value is 0 below $r^*$. 	
\end{lemma}

\begin{proof}
We have already seen that this equation has at most one zero on (0,1) so we now just need to show that the function $\mathfrak{G}$ is positive somewhere to have established that it has a zero. Of course, for $r=\vValue^{-1}(0)$, the function is negative.

$\bullet$ \textbf{$\mathbf{G}$ not locally constant near 0}
Since $G$ is a cdf, and hence a non-decreasing function, the first $k$  such that $g^{(k)}(0)\neq 0$ has $g^{(k)}(0)>0$. Otherwise, $G$ would be decreasing around 0. We treat the case where $G$ is constant near 0 later so we now assume that $k$ exists and is finite.

Let $r$ be such that $1-F(r)=\eps$ very small (e.g. $10^{-6}$). Let $c<r$ be such that $(1-F(r))/(1-F(c))<\eta$ very small (e.g. $10^{-3}$) and $\vValue_X(c)>0$. Hence we can use a Taylor approximation to get that 
$$
G\left(\frac{r(1-F(r))}{1-F(x)}\right)\indicator{x\leq c}\simeq g^{(k)}(0)\frac{(r(1-F(r)))^k}{(1-F(x))^{k}} \indicator{x\leq c}\;.
$$
Integrating this out (ignoring at this point possible integration questions), we get 
$$
\Exp{\vValue_X(X)G\left(\frac{r(1-F(r))}{1-F(X)}\right)\indicator{X\leq c}}\simeq 
g^{(k)}(0)(r(1-F(r)))^k \Exp{\frac{\vValue_X(X)}{[1-F(X)]^k}\indicator{X\leq c}}\;.
$$
Now integration by parts shows that, if $k>1$ 
\begin{gather*}
\Exp{\frac{\vValue_X(X)}{[1-F(X)]^k}\indicator{X\leq c}}=\int_0^c x\frac{f(x)}{(1-F(x))^k}-\frac{1}{(1-F(x))^{k-1}}dx\\
=\left.\frac{1}{k-1}\frac{x}{(1-F(x))^{k-1}}\right|_0^c-\left(1+\frac{1}{k-1}\right)\int_0^c\frac{dx}{[1-F(x)]^{k-1}}\;.
\end{gather*}
If $k=1$, using the fact that $(\ln(1-F(x))'=-f(x)/(1-F(x))$, we have 
\begin{gather*}
\Exp{\frac{\vValue_X(X)}{1-F(X)}\indicator{X\leq c}}=\int_0^c x\frac{f(x)}{1-F(x)}-1 dx\\
=\left.-x\ln(1-F(x))\right|_0^c- c+\int_0^c \ln(1-F(x))dx = -c\ln(1-F(c))-c+\int_0^c \ln(1-F(x))dx
\end{gather*}

We now assume that $k>1$; the adjustments for $k=1$ are trivial and are left to the reader. Clearly, when $F(c)$ is close to 1, we have, since we assume that $f(c)\neq 0$ and $f$ is continuous near 1,  
$$
\int_0^c\frac{dx}{[1-F(x)]^{k-1}}\sim_{c \tendsto 1} \frac{1}{f(c)}\int_0^c \frac{f(x)}{(1-F(x))^{k-1}} dx=\frac{1}{(k-2)f(c)}\frac{1}{(1-F(c))^{k-2}} \;.
$$
So we have, as $c$ increases so that $F(c)\simeq 1$ (while of course having $(1-F(r))/(1-F(c))<\eta$),
$$
\frac{1}{r(1-F(r))}\Exp{\vValue_X(X)G\left(\frac{r(1-F(r))}{1-F(x)}\right)\indicator{X\leq c}}\sim 
[r(1-F(r))]^{k-1}\, g^{(k)}(0)\, \frac{c}{(1-F(c))^{k-1}}>0\;.
$$

Now as $c$ is sufficiently large that we have crossed $\vValue^{-1}(0)$, we have $\vValue(x)> 0$ on $[c,r]$. Of course that implies that $\rho(x)=x(1-F(x))$ is a decreasing function on $[c,r]$.  So we have 
\begin{gather*}
\Exp{\vValue_X(X)G\left(\frac{r(1-F(r))}{1-F(X)}\right)\indicator{c\leq X\leq r}}\geq G\left(\frac{r(1-F(r))}{1-F(c)}\right)\Exp{\vValue_X(X)\indicator{c\leq X\leq r}}\\
=
G\left(\frac{r(1-F(r))}{1-F(c)}\right)(c(1-F(c))-r(1-F(r)))
%\simeq G\left(\frac{r(1-F(r))}{1-F(c)}\right)(r-c) rf(r)
\;.
\end{gather*}
% Now we note that using a Taylor expansion of $1-F(x)$ around $r$, we have
% $$
% x\simeq r+ \frac{F(x)-F(r)}{f(r)}\;.
% $$
So we see that 
\begin{gather*}
\frac{1}{r(1-F(r))}
\Exp{\vValue_X(X)G\left(\frac{r(1-F(r))}{1-F(X)}\right)\indicator{c\leq X\leq r}}\\
\geq 
%rG\left(\frac{r(1-F(r))}{1-F(c)}\right) \frac{F(r)-F(c)}{1-F(r)}
G\left(\frac{r(1-F(r))}{1-F(c)}\right) \left(\frac{c(1-F(c))}{r(1-F(r))}-1\right)\;.
\end{gather*}
If now we take $c_2$ such that $1-F(r)/(1-F(c_2))=1/3$, the reasoning above applies and we have 
$$
\frac{1}{r(1-F(r))}
\Exp{\vValue_X(X)G\left(\frac{r(1-F(r))}{1-F(X)}\right)\indicator{c_2\leq X\leq r}}
\geq  \left(\frac{c_2(1-F(c_2))}{r(1-F(r))}-1\right)G\left(r/3\right) >0\;.
$$
In the last inequality we have used the fact that $c_2<r$ and $\vValue(x)$ is positive so $c_2(1-F(c_2))>r(1-F(r))$. 
Because $\pi_r(x)=(1-F(r))/(1-F(x))$ is increasing, we have $c\leq c_2$, since $\eta=\pi(c)\leq \pi(c_2)=1/3$ . So  we have 
$$
\frac{1}{r(1-F(r))}
\Exp{\vValue_X(X)G\left(\frac{r(1-F(r))}{1-F(X)}\right)\indicator{c\leq X\leq c_2}}\geq 0\;.
$$
Of course the choice of 1/3 above is arbitrary and it could be replaced by any fixed number $s<1$ such that $G(sr)>0$.
We conclude that $\mathfrak{G}$ is positive in a neighborhood of $1$. \\
$\bullet$ \textbf{$\mathbf{G}$ locally constant near 0}
In this case we can pick $c$ such that 
$$
\Exp{\vValue_X(X)G\left(\frac{r(1-F(r))}{1-F(x)}\right)\indicator{X\leq c}}=0\;.
$$
If $c$ is such that $\vValue(c)>0$ our arguments above immediately carry through. 
In fact we can ensure that this is always true by picking such a $c$ and picking a corresponding $r$ as function of the ratio $(1-F(r))/(1-F(c))$ we would like. 

% On the other hand, the rest of the argument carries through and we still have
% $$
% \frac{1}{r(1-F(r))}
% \Exp{\vValue_X(X)G\left(\frac{r(1-F(r))}{1-F(X)}\right)\indicator{c_2\leq X\leq r}}
% \geq  rG\left(r/3\right) >0\;.
% $$
So we conclude that even in this case, 
$\mathfrak{G}$ is positive in a neighborhood of $1$
\end{proof} 

\textbf{Technical note~: } Lemmas \ref{lemma:atMostOneNonTrivialSoln} and \ref{lemma:uniqueSolnInOneStrategicCase} can be extended to the case of distributions that are supported on $[a,\infty]$, where e.g. $a=0$ provided the following properties are satisfied: 
\begin{enumerate}
\item the virtual value crosses 0 exactly once and is negative to the left of the crossing point and positive to the right. 
\item $x(1-F(x))\tendsto 0$ as $x\tendsto \infty$
\item if one can show that $x/(1-F(x))^{k-1}\geq k \int_0^x\frac{1}{(1-F(t))^{k-1}} dt $ for large $x$. Note that this is trivially true for Generalized Pareto distributions (where we can compute the integrals exactly) and for the log-normal distribution (using tail approximations). 
\end{enumerate}
All the arguments go through in exactly the same fashion. One detail may need to be adjusted. In the proof of Lemma \ref{lemma:uniqueSolnInOneStrategicCase}, one may want to define $c_2$ by solving the equation $c_2(1-F(c_2))=\delta r (1-F(r))$, $\delta$ given, $\delta>\eta$ but $\delta$ possibly small and not equal to 1/3. Then $c_2\tendsto \infty$ as $r$ does and $r(1-F(r))/(1-F(c_2))=\delta c_2$.

Finally the argument can be simplified a bit when $F$ is assumed to be regular and supported on $[a,\infty)$ by using the fact that for two increasing functions, $\beta,\gamma$, $\int \beta \gamma\geq \int \beta \int \gamma$. Then one can write 
\begin{gather*}
\Exp{\vValue_X(X)G\left(\frac{r(1-F(r))}{1-F(X)}\right)\indicator{X\leq c}}=
\Exp{\vValue_X(X)G\left(\frac{r(1-F(r))}{1-F(X)}\right)\indicator{X\leq \vValue^{-1}(0)}}\\
+\Exp{\vValue_X(X)G\left(\frac{r(1-F(r))}{1-F(X)}\right)\indicator{\vValue^{-1}(0)\leq X\leq c}}
\end{gather*}
A little algebra, using the fact that $\vValue_X$ and $G(r(1-F(r))/(1-F(x)))$ are increasing show that we can lower bound the second term and using a coarse bound on the first term (bounding it by the minimum value under the expectation) we can finally show that the expectation of interest is positive for $c$ sufficiently large.

\begin{theorem}
We consider the symmetric case and assume that bidders have a compactly supported and regular distribution. We assume for simplicity that the distribution is supported on [0,1]. 

Suppose this distribution has density that is continuous at 0 and 1 with $f(1)\neq 0$. 

Then there is a unique symmetric equilibrium in the class of shading functions defined in Definition \ref{definition_thresholded_strategies}. 

It is found by solving Equation \eqref{app:eq:keyEqThreshStratAllStrategicReStatement} to determine $r^*$ and bidding truthfully beyond $r^*$.
	
\end{theorem}

\begin{proof}
We already know that there is at most one solution since Equation \eqref{app:eq:keyEqThreshStratAllStrategicReStatement} has exactly one solution. 	

If all the bidders but one put themselves at this strategy, we know from Lemma \ref{lemma:uniqueSolnInOneStrategicCase}, which applies because of our assumptions on $f$, that the optimal strategy for bidder one is unique in the class we consider and consists in using a shading that is truthful beyond $r$. This $r$ is uniquely determined by Equation \eqref{app:eq:keyEqThreshStratOneStrategic} but given the shading used by the other players we know that the $r$ determined by Equation \eqref{app:eq:keyEqThreshStratAllStrategicReStatement} is a solution. Hence we have an equilibrium. 
\end{proof} 
\textbf{Uniform[0,1] example} When $K=2$, the solution of Equation \eqref{app:eq:keyEqThreshStratAllStrategicReStatement} and hence the equilibrium is obtained at $r=3/4$. For $K=3$, $r=2/3$; $K=4$ gives $r=15/24=0.625$; $K=5$ gives $r=.6$.

% We recover some of the results of \cite{tang2016manipulate} but with a totally different approach. In their case, since they are using a non-constructive proof they are not able to exhibit the strategies corresponding to the Bayes-Nash equilibrium.

With appropriate shading functions, the bidders can recover the utility they would get when the seller was not optimizing her mechanism to maximize her revenue. Nevertheless, this equilibrium can be weakly collusive since we restrict ourselves to the class of functions introduced in Definition \ref{definition_thresholded_strategies}. It is not obvious that the strategy exhibited in Theorem \ref{thm:revBuyerInThreshStrategy} is an equilibrium in a larger class of functions, because our results in Lemma \ref{lemma:thresholdingIsABestResponse} just imply that thresholding is one possible best response but not the unique one. However, as mentioned previously, from a practical standpoint, as of now there exists no other clear way to increase drastically bidders utility that is independent from a precise estimation of the competition. The fact that at symmetric equilibrium bidders recover the same utility as in a second price auction with no reserves arguably makes it  an even more natural class of bidding strategies to consider from the bidder standpoint.

\subsection{Equivalence of revenue}

\begin{theorem}
Suppose we are in a symmetric situation and all buyers use the symmetric optimal strategy described in Lemma \ref{lemma:uniqueSolnInOneStrategicCase}.

Then the revenue of the seller is the same as in a second price auction with no reserve price. The same is true of the revenue of the buyers. 	
	
\end{theorem}
Interestingly, the theorem shows that this shading strategy completely cancels the effect of the reserve price. This is a result akin to our result on the Myerson auction. 

\begin{proof}
The revenue of the seller per buyer is 
$$
\Exp{\vValue_X(X)F^{K-1}(X)\indicator{X\geq r^*}}, \text{ with }\mathfrak{R}(r^*)=0\;.
$$	
Hence it is also 
$$
\Exp{\vValue_X(X)F^{K-1}(X)\indicator{X\geq r^*}}+\mathfrak{R}(r^*)=\Exp{\vValue_X(X)F^{K-1}(X)}\;.
$$
This is exactly the revenue of the seller in a second price auction with no reserve price. 

From the buyer standoint, his/her revenue is, since all buyers are using the same increasing strategy and the reserve value has been sent to 0, 
$$
\Exp{(X-\vValue_B(\beta(X;r^*)))F^{K-1}(X)}=\Exp{X F^{K-1}(X)\indicator{X\leq r^*}}+
\Exp{(X-\vValue_X(X)) F^{K-1}(X)\indicator{X> r^*}}\;.
$$
We know however that $\mathfrak{R}(r^*)=0$ and therefore
$$
\Exp{X F^{K-1}(X)\indicator{X\leq r^*}}=\Exp{(X-\vValue_X(X)) F^{K-1}(X)\indicator{X\leq r^*}}\;.
$$
Summing things up we get that his/her payoff
$$
\Exp{(X-\vValue_X(X)) F^{K-1}(X)}
$$
as in a second price auction with no reserve. 
\end{proof}

\section{Proof of Section 5: Robustness to the approximation error of the seller}
\label{appendix_section5}
\begin{lemma}
Suppose that the buyer uses a strategy $\beta$ under her value distribution $F$. Suppose the seller thinks that the value distribution of the buyer is $G$. 
Call $\lambda_F$ and $\lambda_G$ the hazard rate functions of the two distributions
Then the seller computes the virtual value function of the buyer under $G$, denoted $\psi_{B,G}$, as 
$$
\psi_{B,G}(\beta(x))=\psi_{B,F}(\beta(x))-\beta'(x)\left(\frac{1}{\lambda_G(x)}-\frac{1}{\lambda_F(x)}\right)\;.
$$	
\end{lemma}
\begin{proof}
% Suppose that the shading strategy $\beta$ belongs to the class defined above; in other words, it has been designed so that below $r$, we have $\psi_B(\beta(x))=0$ under the buyer distribution $F$. To make this dependence on the distribution explicit, we use, in this proof only the notation $\psi_{B,F}$.
As we have seen before we have 
$$
\psi_{B,F}(\beta(x))=\beta(x)-\beta'(x)\frac{1-F(x)}{f(x)}\;.
$$
By construction, we have 
$$
\beta(x)-\beta'(x)\frac{1-F(x)}{f(x)}=0 \text{ for } x\leq r\;.
$$
If the seller perceives the behavior of the buyer under the distribution $G$, we have 
$$
\psi_{B,G}(\beta(x))=\beta(x)-\beta'(x)\frac{1-G(x)}{g(x)}\;.
$$
Hence, we have 
$$
\left|\psi_{B,G}(\beta(x))-\psi_{B,F}(\beta(x))\right|=\left|\beta'(x)\right|
\left|\frac{1-F(x)}{f(x)}-\frac{1-G(x)}{g(x)}\right|\;.
$$
Recall the hazard function $\lambda_F(x)=f(x)/(1-F(x))$. With this notation, we simply have 
$$
\left|\psi_{B,G}(\beta(x))-\psi_{B,F}(\beta(x))\right|=\left|\beta'(x)\right|\left|\frac{1}{\lambda_F(x)}-\frac{1}{\lambda_G(x)}\right|\;.
$$
The corollary follows by noting that when $x\leq r$, 
$$
\left|\beta_{r,\eps}'(x)\frac{1-F(x)}{f(x)}\right|\leq (r-\eps)\frac{1-F(r)}{1-F(x)}
$$
\end{proof}
Hence, a natural way to quantify the proximity of distributions in this context is of course in terms of their virtual value functions. Furthermore, if the buyer uses a shading function such that, under her strategy and with her value distribution, the perceived virtual value is positive, as long as the seller computes the virtual value using a nearby distribution, she will also perceive a positive virtual value and hence have no incentive to put a reserve price above the lowest bid.  In particular, if $\delta$ comes from an approximation error that the buyer can predict or measure, she can also adjust her $\epsilon$ so as to make sure that the seller perceives a positive virtual value for all $x$. 
\subsection{Proof of Theorem \ref{thm:epsAndMonopolyPriceByERM}}
\label{app:subsec:proofERMRobustness}
\begin{customthm}{4}
Suppose the buyer has a continuous and increasing value distribution $F$, supported on $[0,b]$, $b\leq \infty$, with the property that if $r\geq x_2\geq x_1$, $F(x_2)-F(x_1)\geq \gamma_F (x_2-x_1)$, where $\gamma_F>0$. Suppose finally that $\sup_{t\geq r}t(1-F(t))=r(1-F(r))$. 

Suppose the buyer uses the strategy $\newThreshNotation$ described above and samples $n$ values $\{x_i\}_{i=1}^n$ i.i.d according to the distribution $F$ and bids accordingly in second price auctions. Call $x_{(n)}=\max_{1\leq i \leq n}x_i$. In this case the (population) reserve value $x^*$ is equal to 0.

Assume that the seller uses empirical risk minimization to determine the monopoly price in a (lazy) second price auction, using these $n$ samples. 

Call $\hat{x}_n^*$ the reserve value determined by the seller using ERM. 

We have, if $C_n(\delta)=n^{-1/2}\sqrt{\log(2/\delta)/2}$ and $\epsilon>x_{(n)} C_n(\delta)/F(r)$ with probability at least $1-\delta_1$, 
$$
\hat{x}_n^*<\frac{2rC_n(\delta)}{\epsilon\gamma_F} \text{ with probability at least } 1-(\delta+\delta_1) \;.
$$

In particular, if $\epsilon$ is replaced by a sequence $\epsilon_n$ such that \\$n^{1/2}\epsilon_n min(1,1/x_{(n)}) \rightarrow \infty$ in probability, 
$\hat{x}_n^*$ goes to 0 in probability like $n^{-1/2}\max(1,x_{(n)})/\epsilon_n$. 
\end{customthm}
\textbf{Examples : } Our theorem applies for value distributions that are bounded,  with $\epsilon_n$ of order $n^{-1/2+\eta}$, $\eta>0$ and fixed. If the value distribution is log-normal$(\mu,\sigma)$ truncated away from 0 so all values are greater than a very small threshold $t$, standard results on the maximum of i.i.d $\mathcal{N}(\mu,\sigma)$ random variables guarantee that $x_{(n)}\leq \exp(\mu+\sigma \sqrt{2\log(n)})$ with probability going to 1. In that case too, picking $\epsilon_n$ of order $n^{-1/2+\eta}$, $\eta>0$ and fixed, guarantees that the reserve value computed by the seller by ERM will converge to the population reserve value, which is of course 0. 

\textbf{Comment : } The requirement on $\gamma_F$, which essentially means that the density $f$ is bounded away from 0 could also be weakened with more technical work to make this requirement hold only around 0, at least for the convergence in probability result. Similarly one could handle situations, like the log-normal case, where $\gamma_F$ is close to 0 at 0 by refining slightly the first part of the argument given in the proof.
\begin{proof}

$\bullet$ \textbf{Preliminaries}\\
{\sf Notations~: }We use the standard notation for order statistics $b_{(1)}\leq b_{(2)}\leq \ldots\leq b_{(n)}$ to denote our $n$ increasingly ordered bids. 
We denote as usual by $\hat{F}_n$ the empirical cumulative distribution function obtained from a sample of $n$ i.i.d observations drawn from a population distribution $F$. 

Setting the monopoly price by ERM amounts to finding, if $\hat{B}_n$ is the empirical cdf of the bids, 
$$
b^*_n=\argmax_t t (1-\hat{B}_n(t))
$$
We note in particular than 
$$
b_n^*\leq \max_{1\leq i\leq n}b_i=b_{(n)}\;,
$$
since $(1-\hat{B}_n(t))=0$ for $t> b_{(n)}$. 

Because $(1-\hat{B}_n(t))$ is piecewise constant and the function $t\mapsto t$ is increasing, on $[b_{(i)},b_{(i+1)})$ the function $t(1-\hat{B}_n(t))$ reaches its supremum at $b_{(i+1)}^-$. 
$$
b_n^*=\argmax_t t (1-\hat{B}_n(t))=\argmax_{1\leq i \leq n-1} b_{(i+1)}^- \left(1-\frac{i}{n}\right)\;.
$$
Since our shading function $\beta_{r,\eps}$ is increasing and if $x_{(i)}$ are our ordered values, we have, if $\hat{F}_n$ is the empirical cdf of our value distribution,  
$$
\argmax_{1\leq i \leq n-1} b_{(i+1)}^- \left(1-\frac{i}{n}\right)=\argmax_{1\leq i \leq n-1} \beta_{r,\eps}(u_{(i+1)}^-) \left(1-\frac{i}{n}\right)=\argmax_u \beta_{r,\eps}(u)(1-\hat{F}_n(u))\;.
$$
The last equality comes again from the fact that $(1-\hat{F}_n(u))$ is piecewise constant and $\beta_{r,\eps}(u)$ is increasing. So in our analysis we can act as if the seller had perfect information of our shading function $\beta_{r,\eps}$.

In what follows we focus on reserve values and denote 
\begin{gather*}
\hat{x}_{r,n}^*=\argmax_{0\leq x\leq r} \beta_{r,\eps}(x)(1-\hat{F}_n(x))\;,
\hat{x}_n^*=\argmax_{0\leq x} \beta_{r,\eps}(x)(1-\hat{F}_n(x))\;,\\
x^*=\argmax \beta_{r,\eps}(x)(1-F(x))
\end{gather*}
The arguments we gave above imply that $\hat{x}_{r,n}^*\leq x_{(n)}$. We will otherwise study the continuous version of the problem. We also note that by construction, $x^*=0$, though we keep it in the proof as it makes it clearer. 

% Now because $(1-\hat{F}_n(u))$ is piecewise constant and $\beta$ is increasing by construction, we see that on $[x_{j-1},x_j)$, the
% $$
% max_{u\in [x_{j-1},x_j)}=\beta(x_j^{-})(1-\hat{F}_n(x_j^{-}))\;.
% $$
% Hence
% $$
% \max_u \beta(u)(1-\hat{F}_n(u))=\max_{1\leq j \leq n}\beta(x_j^{-})(1-\hat{F}_n(x_j^{-}))\;.
% $$
% 

We recall one main result of \cite{MassartTightConstantDKWAoP90} on the Dvoretzky-Kiefer-Wolfowitz (DKW) inequality: if $C_n(\delta)=n^{-1/2}\sqrt{\log(2/\delta)/2}$, 
$$
P(\sup_x |\hat{F}_n(x)-F(x)|>C_n(\delta))\leq \delta\;.
$$
In what follows, we therefore assume that we have a uniform approximation 
$$
\forall x\;, \; |\hat{F}_n(x)-F(x)|\leq C_n(\delta)\;,
$$
since it holds with probability $1-\delta$. In what follows we write $C_n$ instead of $C_n(\delta)$ for the sake of clarity. Using the fact that $\beta_{r,\eps}$ is increasing, this immediately implies that with probability at least $1-\delta$, for any $c>0$
$$
\forall x \in [0,c]\;, \; |\beta_{r,\eps}(x)(1-\hat{F}_n(x))-\beta_{r,\eps}(x)(1-F(x))|\leq \beta_{r,\eps}(c) C_n\;.
$$

% Of course later we will use the DKW inequality to get $C$ of order $n^{-1/2}$ with high proba.
% We claim that $\argmax \beta(x_j^{-})(1-\hat{F}_n(x_j^{-}))$ cannot be too far from $\argmax \beta(x)(1-F(x))$, which by construction for $x$ is 0.
$\bullet$ {\sf$\hat{x}_{r,n}^*=\argmax_{y\leq r} \beta_{r,\eps}(y)(1-\hat{F}_n(y))$ cannot be too far from $\mathsf{x^*}$}\\
Now for our construction of $\beta_{r,\eps}(x)$, we have by construction that
$$
\frac{\partial }{\partial u}\left[\beta(u) (1-F(u))\right]=-\eps f(u) \text{ when } x\leq r\;.
$$
In particular, it means that when $x,y\leq r$
$$
\beta_{r,\eps}(x)(1-F(x))-\beta_{r,\eps}(y)(1-F(y))=-\eps (F(x)-F(y))\;.
$$
Also $x^*=0$ since $\beta_{r,\eps}(1-F)$ is decreasing on $[0,r]$, as we have just seen that its derivative is negative.  Here we used the fact that $F$ is increasing. 

If $r>y\geq x^*+tC_n/\eps$, we have, using the previous inequality and the fact that  $\beta_{r,\eps}(1-F)$ is decreasing on $[0,r]$,
$$
\beta_{r,\eps}(y)(1-F(y))\leq \beta_{r,\eps}(x^*)(1-F(x^*))-\eps (F(x^*+tC_n/\eps)-F(x^*))\;.
$$
Since we assumed that $F(x_2)-F(x_1)\geq \gamma_F (x_2-x_1)$, we have 
$$
-\eps (F(x^*+tC_n/\eps)-F(x^*))\leq -tC_n \gamma_F\;.
$$
Since $\beta_{r,\eps}$ is increasing, we have $\sup_{0\leq x\leq r}\beta_{r,\eps}(x)\leq \beta_{r,\eps}(r)=r$. 
Picking $t>2r/\gamma_F$, it is clear that for $r>y\geq x+tC_n/\eps$, 
\begin{gather*}
\max_{r\geq u\geq x+tC_n/\eps} \beta_{r,\eps}(u)(1-\hat{F}_n(u))\leq 
\max_{r\geq u\geq x+tC_n/\eps} \beta_{r,\eps}(u)(1-F(u))+ \max_{r\geq u\geq x+tC_n/\eps} \beta_{r,\eps}(u) C_n
\\
\leq 
\beta_{r,\eps}(x^*)(1-F(x^*))+(r-t\gamma_F)C_n< \beta_{r,\eps}(x^*)(1-F(x^*))-rC_n\leq \beta_{r,\eps}(x^*)(1-\hat{F}_n(x^*))\;.
\end{gather*}
We conclude that $\hat{x}_{r,n}^*$ cannot be greater than $x+2rC_n/(\eps\gamma_F)$ and therefore
$$
\hat{x}_{r,n}^*-\hat{x}<\frac{2rC_n}{\eps\gamma_F}. 
$$
$\bullet$ \textbf{Dealing with $\max_{y>r} \beta_{r,\eps}(y)(1-\hat{F}_n(y))$}\\
Recall that $\max_{x} \beta_{r,\eps}(x)(1-F(x))=\beta_{r,\eps}(0)(1-F(0))=r(1-F(r))+\eps F(r)$. 
We now assume that $\max_{y\geq r} y(1-F(y))=r(1-F(r))$. This is in particular the case for regular distributions, which are commonly assumed in auction theory. 

To show that the argmax cannot be in $[r,b]$ with pre-specified probability we simply show that the estimated value of the seller revenue at reserve value 0 is higher than $\max_{y>r} \beta_{r,\eps}(y)(1-\hat{F}_n(y))$. Of course, 
$$
\max_{y>r} \beta_{r,\eps}(y)(1-\hat{F}_n(y))=\max_{x_{(n)}\geq y>r} \beta_{r,\eps}(y)(1-\hat{F}_n(y))\;.
$$

Recall that $\beta_{r,\eps}(0)=r(1-F(r))+\eps F(r)$. Under our assumptions, we have 
\begin{gather*}
\beta_{r,\eps}(0)(1-\hat{F}_n(0))=\beta_{r,\eps}(0)=r(1-F(r))+\eps F(r) \text{ and }\\
\max_{x_{(n)}\geq y\geq r} \beta_{r,\eps}(y)(1-\hat{F}_n(y))\\
\leq \max_{x_{(n)}\geq y\geq r} \beta_{r,\eps}(y)(1-F(y))+C_n \max_{x_{(n)}\geq y\geq r}\beta_{r,\eps}(y)\leq r(1-F(r))+C_n x_{(n)}
\end{gather*}
So as long as $\eps> x_{(n)}C_n/F(r)$, the result we seek holds. By assumption this property holds with probability $1-\delta_1$. 

The statement of the theorem holds when both parts of the proof hold. Since they hold with probability at least $1-\delta$ and $1-\delta_1$ the intersection event holds with probability at least $1-\delta-\delta_1$, as announced.  
\end{proof}

$\bullet$ \textbf{Asymptotic statement/Convergence in probability issue}\\
This is a straightforward application of the previous result and we give no further details.
\end{document}